\documentclass[acmsmall,screen]{acmart}
\usepackage[utf8]{inputenc} 

\usepackage{geometry} 

\usepackage{graphicx} 

\usepackage{booktabs} 
\usepackage{array} 
\usepackage{paralist} 
\usepackage{verbatim} 
\usepackage{subcaption}
\usepackage{listings}
\usepackage[capitalise]{cleveref}
\crefformat{section}{\S#2#1#3} 
\crefmultiformat{section}{\S#2#1#3}{ and~\S#2#1#3}{, \S#2#1#3}{, and~\S#2#1#3} 
\crefrangeformat{section}{\S#3#1#4 to~\S#5#2#6} 

\AddToHook{env/lemma/begin}{\crefalias{theorem}{lemma}}
\AddToHook{env/corollary/begin}{\crefalias{theorem}{corollary}}

\usepackage{algorithm}

\theoremstyle{remark}

\usepackage{fancyhdr} 
\usepackage[normalem]{ulem} 

\newcommand{\smartparagraph}[1]{\vspace{2pt plus .5pt minus 2pt}\textit{#1.}}

\newcommand{\textdef}[1]{\textbf{\emph{#1}}}

\newcommand{\set}[1]{\{#1\}}
\newcommand{\pset}[2]{\set{\,#1\mid#2\,}}
\newcommand{\nat}{\mathbb{N}}

\newcommand{\procs}{\mathcal{P}}
\newcommand{\proc}[1]{\mathtt{#1}}
\newcommand{\procA}{\proc{p}}
\newcommand{\procB}{\proc{q}}

\newcommand{\client}[1]{\proc{c}_{#1}}
\newcommand{\worker}[1]{\proc{w}_{#1}}
\newcommand{\database}{\proc{d}}

\newcommand{\clients}{\mathcal{C}}
\newcommand{\workers}{\mathcal{W}}
\newcommand{\proj}[2]{{#1}|_{#2}}

\newcommand{\internal}{\tau}

\newcommand{\actions}{A}

\newcommand{\action}{a}
\newcommand{\actionB}{b}
\newcommand{\iactions}{I}
\newcommand{\oactions}{O}
\newcommand{\uactions}{U}
\newcommand{\uaction}{u}
\newcommand{\ordactions}{\actions^{<}}

\newcommand{\states}{S}\renewcommand{\state}{s}
\newcommand{\transitions}{\Delta}
\newcommand{\init}{{\mathsf{init}}}

\newcommand{\exec}{\rho}
\newcommand{\trace}{\mathit{tr}}
\newcommand{\lang}{\mathcal{L}}

\newcommand{\messages}{M}
\newcommand{\msg}{m}
\newcommand{\snd}[3]{#1.#2!#3}
\newcommand{\rcv}[3]{#1.#2?#3}
\newcommand{\ccssnd}[2]{#1!#2}
\newcommand{\ccsrcv}[2]{#1?#2}

\newcommand{\system}{{\mathcal{T}}}
\newcommand{\CLTS}[1]{\ensuremath{\{#1_\procA\}_{\procA \in \procs}}}

\newcommand{\channels}{\ensuremath{\mathsf{Chan}}}
\newcommand{\channel}[2]{\ensuremath{#1,#2}}
\newcommand{\chancontents}{\xi}
\newcommand{\lstates}{\mathbf{\state}}
\newcommand{\gstate}{\sigma}

\newcommand{\emptystring}{\varepsilon}

\newcommand{\spec}[1]{T_{#1}}
\newcommand{\clientspec}{\spec{\client{}}}
\newcommand{\workerspec}{\spec{\worker{}}}
\newcommand{\databasespec}{\spec{\database}}

\newcommand{\updatealg}{\mathcal{U}}

\newcommand{\poexec}{\varphi}
\newcommand{\startstate}{\gstate_0}
\newcommand{\porder}{<}
\newcommand{\cut}{C}

\newcommand{\update}{\exec_u}
\newcommand{\pmeasure}{\mathit{pm}}

\newcommand{\rewriteinv}{\mathcal{I}}
\newcommand{\rewrite}{R}

\newcommand{\draftonly}[1]{\ifdefined\finalversion\else#1\fi}
\newcommand{\defineAuthor}[3]{
  \expandafter\newcommand\csname #1\endcsname[1]{%
    \ifdefined\finalversion{##1}%
    \else{\ifdefined\monochrome{\color{green!55!black}{##1}}%
      \else{\color{#3}##1}\fi}%
    \fi}
  \expandafter\newcommand\csname #1comment\endcsname[1]{%
    \ifdefined\finalversion{}%
    \else{\ifdefined\monochrome{}%
      \else{\color{#3}\csname #1\endcsname{#2: ##1}}\fi}%
    \fi}
  \expandafter\newcommand\csname #1out\endcsname[1]{%
    \ifdefined\finalversion{}%
    \else{\ifdefined\monochrome{\color{red!70!black}{\sout{##1}}}%
      \else{\color{#3}{\sout{##1}}}%
      \fi}%
    \fi}
  \expandafter\newcommand\csname #1mout\endcsname[1]{%
    \ifdefined\finalversion{}%
    \else{\ifdefined\monochrome{\color{red!70!black}{\text{\sout{\ensuremath{##1}}}}}%
      \else{\color{#3}{\text{\sout{\ensuremath{##1}}}}}%
      \fi}%
    \fi}
  \expandafter\newcommand\csname #1footnote\endcsname[1]{%
    \ifdefined\finalversion{}%
    \else{\ifdefined\monochrome{}%
      \else{\csname #1\endcsname{\footnote{\csname #1\endcsname{#2: ##1}}}}%
      \fi}%
    \fi}
}
\defineAuthor{tw}{TW}{purple}
\defineAuthor{kn}{KN}{red}
\defineAuthor{dc}{DC}{blue}

\setcopyright{cc}
\setcctype{by-nc-nd}
\acmDOI{10.1145/3776700}
\acmYear{2026}
\acmJournal{PACMPL}
\acmVolume{10}
\acmNumber{POPL}
\acmArticle{58}
\acmMonth{1}
\received{2025-07-10}
\received[accepted]{2025-11-06}

\newtoggle{techrep}
\toggletrue{techrep}
\iftoggle{techrep}{
    \newcommand{\appendixRef}[1]{\cref{#1}}
}{
    \newcommand{\appendixRef}[1]{the extended version~\cite{chaitroth2025consistentupdatesscalablemicroservices}}
}

\newcommand{\finalversion}{}

\title{Consistent Updates for Scalable Microservices}

\author{Devora Chait-Roth}
\orcid{0009-0002-2905-075X}
\affiliation{%
  \institution{New York University}
  \city{Manhattan}
  \country{USA}
}
\email{dc4451@nyu.edu}

\author{Kedar S. Namjoshi}
\orcid{0000-0002-6379-2442}
\affiliation{%
  \institution{Nokia Bell Labs}
  \city{Murray Hill}
  \country{USA}
}
\email{kedar.namjoshi@nokia-bell-labs.com}

\author{Thomas Wies}
\orcid{0000-0003-4051-5968}
\affiliation{%
  \institution{New York University}
  \city{Manhattan}
  \country{USA}
}
\email{wies@cs.nyu.edu}



\begin{document}

\begin{abstract}
  Online services are commonly implemented with a scalable microservice architecture, where isomorphic workers process client requests, recording persistent state in a backend data store. To maintain service,\knout{any} modifications to service functionality must be made on the fly -- i.e., as the service continues to process client requests -- but doing so is challenging. The central difficulty is that of avoiding\knout{potential} inconsistencies \kn{from}\knout{caused by} \emph{mixed-mode} operation, \kn{caused by}\knout{where} workers of current and new versions\knout{are concurrently active and} interact\kn{ing} via the data store. Some update methods avoid mixed-mode altogether, but only at the cost of substantial inefficiency -- by doubling resources (memory and compute), or by halving throughput. The alternative is an\knout{so-called} \kn{uncontrolled} ``rolling'' update, which\knout{is uncontrolled and} runs the risk of serious service failures arising from inconsistent mixed-mode behavior. 

  \dc{Ideally, it should appear to every client that a service update takes effect atomically; this ensures that a client is not exposed to inconsistent mixed-mode behavior. In this paper, we introduce a framework that formalizes this intuition and develop foundational theory for reasoning about update consistency. We apply this theory to derive the first algorithms that guarantee consistency for mixed-mode updates. The algorithms rely on semantic properties of service actions, such as commutativity. We show that this is unavoidable, by proving that any semantically oblivious mixed-mode update method must allow inconsistencies.}

  \dcout{In this paper, we present the first algorithms that guarantee consistency for mixed-mode updates. The algorithms rely on semantic properties of service actions, such as commutativity. We show that semantic awareness is required, by proving that any semantically oblivious, mixed-mode update method cannot avoid inconsistencies. Ideally, it should appear to every client that a service update takes effect atomically; this ensures that a client is not exposed to inconsistent mixed-mode behavior. We introduce a framework that formalizes this intuition and develop foundational theory for reasoning about the consistency of mixed-mode updates, applying that theory to derive the new algorithms and establish their correctness.}
\end{abstract}

\maketitle

\section{Introduction}
\label{sec:introduction}

Online software services, such as services for email, chat, travel reservations, and the like, are expected to be always available. This requirement prevents such services from being shut down to modify their functionality: the familiar shutdown-update-reboot cycle that one uses to update a personal computer is inapplicable. 
Service functionality must instead be updated \emph{on the fly}, i.e., as the service continues to process client requests. 
This work studies on-the-fly updates for the scalable microservice architecture that is common for online services. In a scalable microservice, multiple isomorphic workers process client requests, recording persistent data in a backend data store. This architecture makes it easy to scale the service up or down by controlling the number of active workers.
On-the-fly updates are accomplished by shifting requests from non-updated to updated workers without stopping service; precisely how this shift occurs determines performance and functionality during an update.

A ``blue/green'' update \dc{\cite{blue-green-martin-fowler}} first builds a full system replica containing the new version. Once this replica is running smoothly, traffic is switched to the replica, and the original system is shut down. But replicating a large system is expensive, difficult, and time-consuming. A ``big flip'' update \dc{\cite{brewer-giant-scale-services}} avoids replication at the cost of throughput. It shuts down half the workers and updates those, then shuts down the other half, bringing back the original set of (now updated) workers, before updating and bringing back the remainder. Hence the service offers only half of its regular throughput during an update, which may lengthen client request queues and increase the latency of servicing requests.  

Why deploy such inefficient updates? The motivation lies in avoiding \emph{mixed-mode} operation: if updated and non-updated workers are concurrently active, they can interact indirectly through the data store, leading to unexpected service behavior and possibly to serious failures. 
Such risk is taken by a ``rolling'' update \dc{\cite{brewer-giant-scale-services}}, which updates a small number of workers at a time while all other workers remain available. Resources are conserved and throughput reduction is minimal, but, as the service operates in an uncontrolled mixed-mode during the update, the risks are great.
A spectacular example is that of Knight Capital, a major financial trading company, that went bankrupt within 45 minutes in 2012 due to a mixed-mode error: new-version messages were incorrectly routed to an old-version worker, triggering an infinite cycle of trades \cite{knight-capital-sec}. 
A recent paper in SOSP 2021~\cite{survey:conf/sosp/ZhangYJSRLY} studies 120 distributed systems update failures over the span of 10 years. The authors find that the \emph{majority} of errors were due to incompatible version interactions, and that most of the errors studied were catastrophic.

The challenge of achieving a consistent, efficient, on-the-fly update is thus unresolved. The state of the art is unsatisfactory: one must either accept substantial inefficiencies with blue/green or big-flip updates, or run the risk of serious failures arising from mixed-mode inconsistency in a rolling update. 
Surprisingly, precise formulations of update consistency and formal reasoning about mixed-mode updates are largely lacking from the literature, precluding the design of provably consistent rolling updates. This work aims to fill the gap. We propose a new and (we believe) natural formulation of update consistency, develop methods for reasoning about the consistency of mixed-mode updates, and use those methods to prove the consistency of new mixed-mode update algorithms.

We begin by proposing a formulation for update consistency. The intuition is that external clients of a service should not be aware of how an update is carried out; in particular, whether it involves mixed-mode operation. Our formulation thus views an update computation as being \emph{consistent} if, from the viewpoint of each client, the service \emph{appears} to update \emph{atomically} from its current to the new version at a globally-quiescent state. A quiescent atomic update, by definition, does not exhibit mixed-mode behavior, so this formulation ensures that any internal mixed-mode operation does not result in behavior that is visible to a client.


\dc{To prove update consistency, we introduce an induction principle that relies on rewriting an update computation incrementally, preserving observational equivalence, until a computation with an atomic update is reached. The key is defining a notion of ``incremental''; we give a progress measure to quantify how close a computation is to an atomic update.

From this formal framework, we derive the first update algorithms to guarantee update consistency in mixed-mode operation, and formally establish their consistency. These algorithms rely on semantic properties of actions, such as commutativity, which allow any update computation to be systematically rewritten to an atomic update.}
\dcout{The conditions rely on semantic properties of actions, such as commutativity. We show that one can combine various sufficiency conditions to guarantee consistency for a wider class of updates.}

We expect the semantic properties of actions to be analyzed prior to an update. However, an update algorithm must \dcout{determine a portion of the sufficiency check}\dc{consult relevant semantic properties} at run time. While this incurs overhead, we show that semantic awareness is necessary\knout{for correctness}: we prove an impossibility result, showing that a semantically oblivious update method cannot avoid inconsistent behavior in mixed-mode.

\dcout{From this formal framework, we derive the first update algorithms to guarantee update consistency in mixed-mode operation, and formally establish their consistency.}

To summarize, the contributions of this work include: \kncomment{doesn't like bulleted contribution lists.} \dccomment{I love bulleted contribution lists. Thomas?}
\begin{itemize}
    \item A client-focused formulation of update consistency, 
    \item \dc{An induction principle for proving update consistency,}\dcout{Sufficient conditions that guarantee update consistency for scalable microservices,} 
    \item The first consistent mixed-mode update algorithms for scalable microservices, achieved through semantic awareness, and
    \item A\kn{n impossibility result showing}\knout{proof} that semantic awareness is required for consistent mixed-mode updates. 
\end{itemize}

\subsection{Scope}
Our formulation of update consistency is broadly applicable to general distributed services. However, the update algorithms are specialized to a commonly applied scalable microservice architecture, where client requests are serviced by a set of isomorphic worker nodes operating on a shared data store. (We also use the term "database" in a colloquial sense.) The updates may alter worker behavior arbitrarily but not add new features, as that may require updating clients as well, nor modify the database interface. We assume that only one update is in progress at a time.

\knout{Finally, we} We make several assumptions about the behaviors of the microservice processes. \knout{\dc{We plan to relax our assumptions and apply these foundations towards more complex systems in future work.}}\dccomment{Removed the blatant repetition from the last paragraph, keep this line or remove?} First, we assume that clients (as well as workers) are mutually independent: I.e., there is no direct communication between  clients nor between  workers. Second, we assume that clients are sequential: a client must receive a response to an outstanding request before sending a new request. We similarly assume that workers only service one request at a time. Third, we assume clients treat the service as a black box and are unaware of its internal structure and scale. And fourth, we assume that workers are effectively stateless: they may store temporary state to calculate responses to clients, but that state is not persistent from one request to the next. Connected to this is the \kn{simplifying technical} assumption that every response calculation\knout{requires} accesses the database \kn{(possibly doing nothing)}. \kn{These}\knout{Such} assumptions are met by\knout{many} common services such as email systems, social media platforms, and chat services. 

\dc{In this work we focus on the foundations of update consistency, using a simple system to more clearly demonstrate our formalization and proof method.}
In reality, large-scale distributed services may have a network of interacting microservices and may apply updates concurrently to these microservices. We expect that the foundation developed in this work for consistent single-microservice update will lead to provably consistent algorithms for combinations of microservices, but we leave that analysis to future work.

\subsection{Prior Work}
The consistency of distributed system updates has been studied (intermittently) since at least the early 1980s. Seminal research includes \knout{work on} the Argus system~\cite{bloom-thesis-1983,DBLP:journals/iee/BloomD93}, DYMOS~\cite{DBLP:journals/sigplan/CookL83}, and CONIC~\cite{DBLP:journals/tse/KramerM90}. 
There, update consistency is formulated in terms of type-safe replacement or preservation of a global \kn{service} invariant. While these criteria provide a measure of consistency, they are incomplete as, for instance, it is not specified how liveness properties are treated (e.g., whether pending requests must be fulfilled). Nor do they apply to updates that cause behavioral change and thus do not preserve all invariants. 
Type- and specification-based formulations of consistency are also part of several language-specific formulations of dynamic update for single-server applications (cf.~\cite{DSU:conf/pldi/HicksMN,mutatis-mutandis:conf/toplas/StoyleHBSN,Ginseng:conf/pldi/NeamtiuHSO,co-specs-DSU:conf/vstte/HaydenMHFF}). (We discuss related work more fully in \cref{sec:related-work}.)

The blue/green and big-flip update mechanisms are folklore, although similar methods have been proposed for the Imago system~\cite{imago:conf/middleware/DumitraN} and for concurrent object-based systems~\cite{modular-upgrades-ds:conf/ecoop/AjmaniLS}. As discussed, consistency is obtained at the price of doubling memory and compute, or by halving throughput.

Network \kn{routing} configuration updates face similar challenges, and consistency for network  updates has been the target of much research (cf.~\cite{UpdateSurvey} for a recent survey). A strong consistency criterion called \emph{per packet consistency} is introduced in~\cite{abs-network-updates:conf/sigcomm/ReitblattFRSW}; this requires every network packet to be entirely processed by either the current or the new configuration, never by a mix. 
The \kn{paper} proposes an update method that requires doubling router memory.
A recent "causal" update method~\cite{in-place-network-updates:conf/sigcomm/NamjoshiGS} obtains the same strong consistency guarantee while operating in place (i.e., without additional memory), but the algorithm may drop packets to guarantee \kn{per-packet} consistency. While networks are designed to tolerate packet losses, software systems are less robust, \kn{and hence software update methods must be lossless}. Applying the causal update method to a microservice may also result in a large number of client requests being dropped. 

The consistency formulation and update algorithms developed in this paper avoid these disadvantages. Consistency is based on an external view of the service, not on internal type- or invariant-preservation. The consistency definition requires the update to appear atomic to every client; thus, in determining whether an update would meet service requirements, designers only have to reason about atomic updates, a simpler task. The update algorithms work in place, without additional memory. They are a form of rolling update, which limits throughput loss. The price for these desirable properties is that \kn{the system actions}\knout{abstract request specifications} must be analyzed for commutativity and other semantic properties. We show that this is unavoidable.

\section{\dcout{Example} \dc{Overview}}
\label{sec:overview}

Consider a basic \dc{messaging} service, where the actions available to clients are sending a \dc{message}, checking their inbox, and reading a \dc{message} from their inbox. Every client has an individual inbox in the data store that stores received \dc{messages}. 
Client requests are sent to and serviced by worker nodes. For instance, when Alice requests to send a \dc{message} to Bob, \knout{Alice's}\kn{the} servicing worker adds the content of the \dc{message} to Bob's inbox. Then when Bob requests to check his inbox, \knout{his}\kn{the} servicing worker retrieves \knout{header data from}his inbox entries and sends that data to Bob. 
This \dc{messaging} service is structured as a scalable microservice \dc{(\cref{sec:formal-model})}, allowing workers to be added or removed in accordance with\dcout{client} traffic.

We assume two important properties of our clients. First, clients are independent from each other: they do not know the exact sequences of requests sent by other clients, nor the worker responses other clients receive. Certainly, if database accesses caused by two clients' requests overlap, the actions of one can be \emph{indirectly} observed by another. 
For example, when Alice sends a \dc{message} to Bob, \knout{Bob}\kn{he} \dc{can} observe the \dc{message} written to his inbox and \dcout{can}deduce that Alice had performed a ``send-\dc{message}'' action. 
But, no client \emph{directly} observes the requests and responses of other clients.
The second important client property is that clients are ignorant of the precise actions performed by workers to service their requests. In other words, the internal behavior of the system is a black box to clients. Their observations are restricted to sending requests and receiving responses.

We may want to update the ``send'' action to attach an automated translation to a \dc{message} based on the country the recipient is based in. Consider a \emph{rolling deployment} of this update, where workers are updated one by one with all other workers remaining available. Now suppose that during the update, Amelie sends a \dc{message} from France, in French, to George in England, and Amelie's ``send'' request is serviced by an updated worker. Much to George's delight, he receives a translation of the \dc{message} in English rather than needing to translate from French himself. George then sends a \dc{message} back to Amelie in English, expecting that his \dc{message} will be automatically translated to French\dcout{when he sends it}. However, George's ``send'' request is serviced by a non-updated worker (as may happen in a rolling update), and no French translation is shown,\dcout{instead} offending Amelie with an English \dc{message}. 

This confusing sequence is caused by the uncontrolled mixture of updated and non-updated workers. George first observes behavior consistent with an updated service, then observes a non-updated behavior (e.g., if Alice sends a response that includes his untranslated \dc{message}), uncovering the system's internal mixed-mode operation. While this particular inconsistency may be viewed as benign,  mixed-mode inconsistencies can have serious consequences, as described in \cref{sec:introduction}. Thus, it is necessary to control the processing of requests during a rolling update. 

In this paper, we introduce the first \dc{formal framework for reasoning about update consistency.}\dcout{provably consistent methods for rolling updates.} \dc{We say that an update is \emph{consistent} if}\dcout{These algorithms ensure that} the service behavior\dcout{that is} observed by each client \dc{aligns}\dcout{is consistent} with an \emph{atomic} update to the service, even though in reality the update process allows concurrent operation of updated and non-updated workers.
This formulation of update consistency \dc{(\cref{sec:consistent-updates})} lets us better understand what went wrong in our example.  George observes that an updated action (Amelie's translated \dc{message}) occurs before a non-updated action (his untranslated \dc{message}). From George's perspective, it is impossible to reconsider those events as arising from an atomic service update.

\dc{From this framework, we develop the first provably consistent algorithms for rolling updates.}
Our algorithms exploit the semantic properties of system actions to control and direct client requests such that every client observes an atomic update. Consider again our \dc{messaging} example. Say Alice sends a \dc{message} to Bob via an updated worker, and then Bob sends a \dc{message} to Claire via a non-updated worker. From the perspective of the clients, this trace is indistinguishable from a trace that reverses the order of these operations (since Bob does not check his inbox in between). Both actions write to the database, but regardless of which action is ordered first, they result in the same final database state and client observations. In other words, the two actions \emph{commute}. \kn{Bob may assume that his message to Claire was processed before Alice sent her message, with the system updating atomically in between.}
When \dc{a series of} actions \dc{in a computation} commute, they can be \dc{repeatedly reordered until we} produce a trace with an atomic update \dc{(\cref{sec:comm-update-algorithms})}\dcout{: clients can assume that first Bob sent his non-updated message and then Alice sent her updated message, with the system updating atomically in between}. Since clients are unaware of the actual trace beyond their own observations, the reordered trace is equally plausible to clients\dcout{as the actual one}. It appears to them that the system updated atomically, giving us update consistency.

We can also exploit the semantic properties of \emph{backward-} and \emph{forward-compatibility} to obtain update consistency \dc{(\cref{sec:compat-update-algorithms})}. 
Backward-compatibility colloquially refers to instances when new-version workers can service requests in a manner consistent with a non-updated system; forward-compatibility is analogous for old-version workers. 
Suppose in our \dc{messaging} service that a ``format'' action is available to clients to draft a \dc{message} with special text formatting, like creating a bulleted list. Now suppose that we update the system so that any \dc{message} sent with an asterisk on a new line is automatically formatted into a bulleted list, without requiring the ``format'' action first, unless the asterisk is preceded by a backslash. If Alice has observed an updated system and she writes a \dc{message} with an asterisk on a new line, that request can be serviced by an old-version worker if it is \emph{translated} to an explicit ``format'' request. Then the old-version worker produces results consistent with an updated system: the request is forward-compatible, which can hide the internal mixed-mode operation from clients. Likewise, if Alice is expecting an old-version response, then her \dc{message} with an asterisk on a new line can be sent to a new-version worker if it is translated to a \dc{message} with a backslash; backward-compatibility ensures that the new version worker's response will appear to Alice like that of an old-version worker.

To guarantee consistency, our update algorithms control how client requests are routed to updated or non-updated workers. The commutativity update algorithm (\cref{alg:basic-comm}) enforces that all non-updated worker actions left-commute with all updated worker actions during the update, so that clients may (conceptually) reorder the actual computation to one where all non-updated actions occur before all updated actions. The backward-compatibility update algorithm (\cref{alg:backward-compatible}) only directs backward-compatible translations to updated workers, so that clients only observe the update with the final updated worker; the forward-compatibility algorithm (\cref{alg:forward-compatible}) flips \knout{the} \kn{those} conditions. \cref{alg:comm-update-inv} combines the commutativity and compatibility update strategies. 

\dc{We model system processes as communicating labeled transition systems (\cref{sec:formal-model}), and update algorithms as yet another transition system which restricts the processes' nondeterministic choice of who to communicate with, when to communicate, and when to update (\cref{sec:consistent-updates}).}
The core of this paper presents the algorithms and their consistency proofs \dc{(\cref{sec:comm-update-algorithms}, \cref{sec:proving-update-consistency}, \cref{sec:compat-update-algorithms})}. While the algorithms are simple to state and to implement, their consistency proofs are surprisingly intricate and it is necessary to perform a careful analysis of several subtle cases. 
\dc{Our consistency proofs rely on a partial order view of computations (\cref{sec:proving-update-consistency}). In this view, an atomic update becomes a consistent cut of the partially-ordered computation, dividing cleanly between updated and non-updated events. The correctness proofs show that events in any update computation controlled by the algorithm can be systematically reordered to converge on an atomic update.}
We also prove that any consistent rolling update method must consider the semantics of client requests; any method that is oblivious to semantics will exhibit a mixed-mode inconsistency (\cref{subsec:impossibility-result}).

\section{Formal System Model}
\label{sec:formal-model}

In a \textdef{scalable microservice}, such as the \dc{messaging} service discussed in \cref{sec:overview}, client requests are serviced by a set of interchangeable workers who operate on a shared database. 
Clients operate independently and are oblivious to the internal workings of the service. The number of workers can change during operation as traffic demands, making such microservices scalable. In practice, a service implementation includes a load balancer, which ensures that the load from client requests is distributed across workers in a manner that meets certain criteria (e.g., evenly distributed load). As this concerns performance rather than correctness, we ignore the presence of the load balancer and instead imagine client requests being sent to workers directly.

\subsection{Scalable Microservices}
\label{subsec:scalable-microservices}
\dcout{We treat each process in a microservice (clients, workers, and database) as a reactive labeled transition system, where each process can send messages to, and receive messages from, other processes via one-way FIFO channels.}

We start by defining some basic notions. 
A \textdef{labeled transition system} (LTS) is a tuple
$T = (\states, \actions, \transitions, \state_\init)$
where
$\states$ is a set of states, $\actions$ is a set of action labels,
\knout{$\transitions$ is a set of transitions from $S \times \actions \times \states$,}
\kn{$\transitions \subseteq \states \times \actions \times \states$ is a set of transitions},
and $\state_\init \in \states$ is the initial state. 
We denote a transition $(\state, \action, \state') \in \transitions$ by $\state \xrightarrow{\action} \state'$.

\dcout{An \textdef{execution} of an LTS $T$ is an infinite alternating sequence of states and actions $\exec = \state_0; \action_0; \state_1; \action_1; \dots$ where for each $k$, $\state_k \xrightarrow{\action_k} \state_{k+1}$. The \textdef{trace} of $\rho$ is the sequence of its actions, denoted $\trace(\rho)=\action_0,\action_1,\dots$.
We call $\exec$ a \textdef{computation} if it starts in the initial state, $\state_0=\state_\init$. 
We define a \textdef{computation fragment} as a finite subsequence of a computation, to allow us to examine only a finite portion of a computation. A computation fragment is called initial if it starts in $\state_\init$.
We write $\lang(T)$ for the set of all traces of $T$'s computations.
For $B \subseteq \actions$ and a trace $\alpha$, we denote by $\proj{\alpha}{B}$ the homomorphic projection of $\alpha$ onto $B$ (i.e., the subsequence of $\alpha$ containing actions in $B$).
We lift projection to sets of traces in the expected way.
}

\dc{We model each process in a microservice (clients, workers, and database) as a\knout{reactive} labeled transition system, where each process can send messages to, and receive messages from, other processes via one-way FIFO channels.}
Let $\procs$ be a finite set of processes and $\messages$ a set of message values. For $\procA,\procB \in \procs$ and $\msg \in \messages$, we write $\snd{\procA}{\procB}{\msg}$ for the action of $\procA$ sending $\msg$ to $\procB$ and $\rcv{\procA}{\procB}{\msg}$ for $\procA$ receiving $\msg$ from $\procB$. The set of $\procA$'s \textdef{observable actions} is $\oactions_\procA = \pset{\snd{\procA}{\procB}{\msg},\rcv{\procA}{\procB}{\msg}}{\procB \in \procs \setminus \set{\procA}, \msg \in \messages}$,\knout{Additionally, we assume a set of mutually disjoint internal actions $\internal \in \iactions_\procA$} \kn{along with a disjoint set $\iactions_\procA$ of internal actions}\knout{for each $\procA \in \procs$}. The set of all actions of $\procA$ is thus $\actions_\procA = \oactions_\procA \cup \iactions_\procA$.

A \textdef{communicating transition system} $\mathcal{T}$ over $\procs$ and~$\messages$ is a tuple $(\CLTS{T},\channels)$ consisting of a labeled transition system ${T}_\procA = (\states_\procA, \actions_\procA, \transitions_\procA, \state_{\init, \procA})$ for every $\procA\in\procs$ and the set of channels $\channels \subseteq \set{(\channel{\procA}{\procB}) \mid \procA,\procB\in \procs, \procA\neq \procB}$. \dc{We model a microservice as a communicating transition system over the set of clients, workers, and the database, and over the set of messages (requests, responses, database operations, etc.) sent between them. \cref{subsec:architect} elaborates on the specific architecture we assume.}

A \textdef{global state} $\gstate$ of $\mathcal{T}$ is a pair $(\lstates,\chancontents)$ where $\lstates \in \Pi_{\procA \in \procs}  \states_\procA$ tracks the local state of each process and $\chancontents \in \channels \to \messages^*$ the channel contents.
We denote the projection from a global state $\gstate$ to the local state $\lstates(\procA)$ of an individual process $\procA$ as $\proj{\gstate}{\procA}$.
In the initial state $\gstate_\init=(\lstates_\init, \chancontents_\init)$ of $\mathcal{T}$, each process $\procA$ is in its initial local state, $\lstates_\init(\procA) = s_{\init,\procA}$, and $\chancontents_\init$ maps each channel to the empty sequence $\emptystring$.

Each transition $\gstate \xrightarrow{\action} \gstate'$ of $\mathcal{T}$ has an active process $\procA$ that performs the action $\action \in \actions_\procA$, causing the global state to be updated according to the following rules: 
\begin{itemize}
	\item
	$(\lstates,\chancontents) \xrightarrow{\snd{\procA}{\procB}{\msg}} (\lstates[\procA \mapsto s'],\chancontents[(\channel{\procA}{\procB}) \mapsto \chancontents(\channel{\procA}{\procB}) \cdot \msg])$ if
	$(\lstates(\procA), \snd{\procA}{\procB}{\msg}, s')\in\transitions_\procA$ and $(\channel{\procA}{\procB}) \in \channels$.

	\item $(\lstates,\chancontents) \xrightarrow{\rcv{\procA}{\procB}{\msg}} (\lstates[\procA \mapsto s'],\chancontents[(\channel{\procB}{\procA}) \mapsto w])$ if
	$(\lstates(\procA), \rcv{\procA}{\procB}{\msg}, s')\in\transitions_\procA$ and $\chancontents(\channel{\procB}{\procA}) = \msg \cdot w$.
	
	\item $(\lstates,\chancontents) \xrightarrow{\internal} (\lstates[\procA \mapsto s'],\chancontents)$ if
	$(\lstates(\procA), \internal, s')\in\transitions_\procA$ and $\internal \in \iactions_\procA$.
	
\end{itemize}

\dc{An \textdef{execution} of an LTS $T$ (and similarly, of a communicating transition system $\mathcal{T}$) is an infinite alternating sequence of states and actions $\exec = \state_0; \action_0; \state_1; \action_1; \dots$ where for each $k$, $\state_k \xrightarrow{\action_k} \state_{k+1}$. The \textdef{trace} of $\rho$ is the sequence of its actions, denoted $\trace(\rho)=\action_0,\action_1,\dots$.
We call $\exec$ a \textdef{computation} if it starts in the initial state, $\state_0=\state_\init$. 
We define a \textdef{computation fragment} as a finite segment of a computation.
A computation fragment is called initial if it starts in $\state_\init$.
We write $\lang(T)$ for the set of all traces of $T$'s computations.

For $B \subseteq \actions$ and a trace $\alpha$, we denote by $\proj{\alpha}{B}$ the homomorphic projection of $\alpha$ onto $B$ (i.e., the subsequence of $\alpha$ containing actions in $B$).
\dc{This}\dcout{The homomorphic projection} allows us to project out from a computation the actions of a given set. For example, if $A$ is the set $\{a,b,c\}$ and $B$ is the set $\{b,c\}$, the projection of a trace $\alpha = acabacc$ onto $B$, $\proj{\alpha}{B}$, is $cbcc$. We use projections throughout the paper to obtain the client observations of a given computation.
We lift projection to sets of traces in the expected way.}

\dcout{Executions, traces, etc.\ of $\mathcal{T}$ are defined as for LTS.}

\subsection{Architectural Assumptions}
\label{subsec:architect}

We make certain architectural assumptions about the communicating transition system $\mathcal{T}$ defining a microservice. First, we assume $\procs = \clients \uplus \workers \uplus \{\database\}$ where $\clients$ is the set of clients, $\workers$ the set of workers, and $\database$ the database process.
The communication topology of a microservice is restricted as follows: there exists a channel from every client to every worker, from every worker to the database, from the database to every worker, and from every worker to every client, i.e. $\channels = \clients \times \workers \cup \workers \times \{\database\} \cup \{\database\} \times \workers \cup \workers \times \clients$.
This sets up an architecture where clients communicate with workers, workers operate on the database, and then workers send responses back to clients.
\knout{Notice that there }There are no channels between distinct clients nor between distinct workers: clients are \textdef{independent} from other clients, and workers are \textdef{independent} from other workers.

Processes can keep local state. Workers' state is divided into instruction state, which determines how a worker computes responses to requests, and general local state. 
However, we\knout{ do} make the assumption that workers' general local state \dc{does not persist between service of different requests; such workers are ``stateless'' in common parlance.}\dcout{ is \emph{not necessary} for worker operation: local state may make it more \emph{efficient} for a given worker to service a given request, but any worker \emph{can} service any request, possibly with additional internal actions.}
\knout{This assumption allows us to say that clients have no way of knowing which worker serviced their requests, preserving the idea that clients are oblivious to the internal workings of the service.}

In this work, we are interested in services where clients send requests \textdef{sequentially}.
For example, in a \dc{messaging} service, a client cannot concurrently send and read a \dc{message}; it can only perform those actions in sequence.
For simplicity, we assume that each worker operates on the database when servicing a request. (This can be a \dc{skip}\dcout{no-op}.) 

We can therefore partition the behavior of a system into a set of \textdef{relays}, where each relay consists of a client request sent from the client to the worker, internal worker operations, database access by the worker, and a client response sent from the worker to the client.
For instance, a relay in our \dc{messaging} service might be a client requesting to read a \dc{message}, her servicing worker finding which portion of the database to read, reading and \dc{(temporarily)} storing the relevant portion of the database, and sending that data back to the client.

Now that we have described the basic structure of a microservice, we\dcout{ can} give\dcout{ a little} more detail about the transition systems of the clients, workers, and database.
We use CCS-like notation for describing LTS's $\clientspec$, $\workerspec$, and $\databasespec$ that provide specifications for the actual client, worker, and database transition systems. These specifications are to be understood in terms of observational refinement. \knout{For example, for each client $\client{} \in \clients$,} \kn{That is, for the actual transition system $T_{\client{}}$ of a client $\client{} \in \clients$,} we require $\proj{\lang(T_{\client{}})}{\oactions_{\client{}}} \subseteq \lang(\clientspec(\client{}))$ and similarly for workers and the database LTS. More precisely, the transition system of a process can deviate from its specification by performing auxiliary internal actions and by constraining the non-deterministic choices of message values in the observable send actions of the specification.

\subsubsection*{Clients}
Clients send sequential requests to workers and receive responses from them:
\[ \clientspec = \mu \mathsf{C}.\; \sum_{\worker{} \in \workers, \mathit{req} \in \messages} \ccssnd{\worker{}}{\mathit{req}}.\, \sum_{\mathit{rsp} \in \messages} \ccsrcv{\worker{}}{\mathit{rsp}}.\, \mathsf{C} \enspace.\]
To avoid notational clutter, we leave the fixed active process $\client{}$ implicit in all actions specified by $\clientspec$, writing e.g. $\ccssnd{\worker{}}{\mathit{req}}$ for $\snd{\client{}}{\worker{}}{\mathit{req}}$.

\subsubsection*{Workers}
Workers receive requests from clients, operate on the database, calculate responses, and send responses back to clients:
\[ \workerspec = \mu \mathsf{W}.\; \sum_{\client{} \in \clients, \mathit{req} \in \messages} \ccsrcv{\client{}}{\mathit{req}}.\, \sum_{\mathit{op} \in \messages} \ccssnd{\database{}}{\mathit{op}}. \, \sum_{\mathit{res} \in \messages} \ccsrcv{\database{}}{\mathit{res}}.\, \sum_{\mathit{rsp} \in \messages} \ccssnd{\client{}}{\mathit{rsp}}.\, \mathsf{W} \enspace.\]
We require that all worker LTS are isomorphic up to renaming of internal actions and the worker's identities in observable actions. \dc{This assumption is necessary to allow arbitrary replacement of workers during service.}

\subsubsection*{Database}
The database receives operation requests from workers and performs each operation atomically, changing the state of the database. (This is an abstraction of actual database operation, which may allow multiple concurrent operations, while ensuring atomicity and isolation.)
\[ \databasespec = \mu \mathsf{DB}.\; \sum_{\worker{} \in \workers, \mathit{op} \in \messages} \ccsrcv{\worker{}}{\mathit{op}}. \, \sum_{\mathit{res} \in \messages} \ccssnd{\worker{}}{\mathit{res}}.\, \mathsf{DB} \enspace.\]

For the remainder of the paper, we assume that the scalable microservices under consideration follow the above architectural assumptions. We refer to these simply as \emph{systems}. We fix a system $\mathcal{T}$ for all of the definitions we make throughout the remainder of this section. Two properties of the architecture follow easily from the definitions: (1) every occurrence of an action of $\mathcal{T}$ is part of exactly one relay, and (2) every relay includes exactly one client, one worker, and the database.

\subsection{Client Projection and Equivalence}

Our definition of update consistency is focused on the external observations of clients. The \textdef{$\procA$-projection} of a computation fragment $\exec$ for a process $\procA$\dcout{, written $\proj{\exec}{\procA}$,} \dc{is $\proj{\trace(\exec)}{\actions_\procA}$}, the homomorphic projection of $\trace(\exec)$ onto $\procA$'s actions $\actions_\procA$. For instance, the $\client{}$-projection of a computation fragment in which a client $\client{}$ requests to read a \dc{message} will only contain $\client{}$'s lookup request for the \dc{message} that it sends to a worker and its reception of the worker's response to that request. Similarly, the $\messages$-projection of a sequence of actions \dc{$\alpha$}\dcout{ $\exec$} is the homomorphic projection of \dc{$\alpha$}\dcout{ $\exec$} onto the messages contained in send and receive actions. The $(\procA{},\messages)$-projection is the \kn{sequential} composition of these two projections\kn{; the composition isolates the sequence of messages sent and received by \dc{process}\dcout{ client} $\procA{}$ while hiding the identities of the \dc{particular processes $\procA$ is communicating with}\dcout{ serving workers}.}

Since clients have an obscured view of a service, many computations will be indistinguishable to a client as long as the client receives the same response messages to their requests. We say that two computation fragments $\exec$ and $\exec'$ are \textdef{$\client{}$-equivalent} for client $\client{}$ if they have the same global start state, their $(\client{},\messages)$-projections are equal, and if both $\exec$ and $\exec'$ are finite, they have the same end state. This definition formalizes the idea that clients cannot distinguish the identity of workers.

For example, consider a computation where Alice sends a \dc{message} to Bob; \knout{Bob}\kn{he} checks his inbox and reads that \dc{message}; and then Charles sends a \dc{message} to Alice, which she receives. This computation is Alice-equivalent to one where Alice sends the \dc{message} to Bob, then receives the \dc{message} from Charles. 
\section{Consistent Updates} 
\label{sec:consistent-updates}

We formalize the notion of an update algorithm and establish that there cannot be a consistent rolling update algorithm that is oblivious to the semantic properties of client requests. Throughout this paper, the assumption is that the current and new versions of the service are, by themselves, free of error. The focus is thus solely on errors that may arise from mixed-version interactions \emph{during} the update process. 

\subsection{System Updates}
\label{subsec:system-updates}

We define a \textdef{system update} as replacing one or more system processes with another process that communicates on identical channels as the original.
In this work, we only consider updates to worker processes. A worker update transforms a worker's instruction state and general local state, but does not change channel state nor channel structure, or alter the types of messages sent or received by workers. These simplifications exclude updates that add new features or remove old features, as those require changes to the set of messages.  

We treat worker updates as occurring instantaneously through dedicated internal worker actions, $\uaction \in \uactions_{\worker{}} \subseteq \iactions_{\worker{}}$. An update action simply replaces the worker's instruction state. An update computation updates every worker exactly once and must eventually update all workers. For any computation, the shortest prefix containing all update actions is a computation fragment that we refer to as an  \textdef{update fragment}, or just ``update'' for short.

A \textdef{quiescent update} is an update where workers are updated only in a \emph{wait} state. Formally, a worker state $\state \in \states_{\worker{}}$ is a wait state of $\worker{}$ if $\worker{}$ accepts client requests in $\state$, i.e., there exists a transition $\state \xrightarrow{\rcv{\worker{}}{\client{}}{\msg}} \state'$ for some $\client{}$, $\msg$, and $\state'$. A quiescent update guarantees that every client request is serviced by either a non-updated worker or an updated worker, never by a mix, and that a worker is never updated while servicing a request. We assume that all updates of the system are quiescent.

An \textdef{atomic update} is an update where all update actions are in a single contiguous fragment that contains no non-update actions.
We make the assumption that service requirements are met by an atomic (quiescent) update. This is a reasonable framing, as such an update corresponds to a shutdown-update-restart computation. These updates are thus at the core of our consistency formulation below.

\begin{definition}
    An update in a computation $\exec$ is \textdef{consistent} if, for every client $\client{} \in \clients$, there exists a computation with an atomic update that is $\client{}$-equivalent to $\exec$.
\end{definition}
\knout{Note that an}\kn{An} atomic update is trivially consistent by definition. In a consistent update computation, the view of any client is identical to its view on a (possibly different and possibly client-specific) atomic update computation. From the discussion above, the behavior of a service as viewed by a client is correct (for a service-dependent notion of correctness) in an atomic update. Hence, the service behavior is also correct in a consistent update computation, as no client can distinguish the actual update computation (which need not be atomic) from one where the update is atomic.

Note that in a consistent update, distinct clients may perceive the system as updating at distinct points: it is a $\forall\exists$ property rather than an $\exists\forall$ property. The stronger requirement of a single point of update perceived by all clients is unnecessarily restrictive, since clients are independent. All that matters is that to each client, the update appears to have taken place atomically at some moment, but those moments need not agree for different clients.

Consider again the language translation example from \cref{sec:overview}. In that computation, George reads a \dc{message} processed by an updated worker, and is subsequently serviced by a non-updated worker. This sequence exposes a non-atomic update: first a worker updated (as revealed by the updated send-\dc{message} action), and following that a non-updated worker performed an action before eventually updating. There does not exist a George-equivalent computation with an atomic update, because one cannot reorder these actions visible to George. 

\subsection{Update Algorithms and Impossibility Result} 
\label{subsec:update-algorithms}
\label{subsec:impossibility-result}

\cref{sec:comm-update-algorithms} presents several consistent rolling update algorithms. These algorithms rely on an analysis of the semantic properties of service actions. We show here that this semantic awareness is necessary by presenting an impossibility theorem that establishes that no update algorithm allowing mixed-mode computations can be universally consistent. For this proof, it is necessary to formalize what we mean by ``update algorithm.''

\subsubsection{Update Algorithms}
\label{subsubsec:update-algs}
A distributed mechanism for deploying an update in practice (inspired by \cite{in-place-network-updates:conf/sigcomm/NamjoshiGS}) is to use auxiliary update \emph{shell} processes around each worker and client that control message flow \dc{and track update status}, along with an update manager component to direct the update system-wide. When the update shell receives an instruction from the update manager to update its worker, the worker is shut down and replaced by an updated worker.

Rather than representing the update shells and manager \dc{of an update algorithm} as explicit processes in the formal model $\mathcal{T}$ of a microservice, we treat them abstractly as another labeled transition system that controls some of the non-deterministic choices made by clients, workers, and the database. In particular, the \dc{transition system}\dcout{ update manager} controls which workers a client may send a request to while an update is in progress and the order in which the database processes operations sent by workers, \dc{abstracting the role of an update manager}. \dcout{The update shell}\dc{Moreover, it} controls which internal update action a worker performs and when, \dc{abstracting the role of update shells}.
We also assume a temporary extension of the message set for update-specific communications. We assume that these messages are included in the set of all messages $\messages$.

Formally, an \textdef{update algorithm} for microservice $\mathcal{T}$ is a labeled transition system $\updatealg$ over a set of \textdef{controllable actions} $\actions_\updatealg \subseteq \actions_{\mathcal{T}}$. The controllable actions consist of all observable actions \dc{(i.e., sends and receives)} and all internal worker update actions of $\mathcal{T}$, $\actions_\updatealg = \oactions_{\mathcal{T}} \cup \bigcup_{\worker{} \in \workers} \uactions_{\worker{}}$. We say that a computation $\exec$ of $\mathcal{T}$ is controlled by $\updatealg$ if its trace is consistent with a computation of $\updatealg$, $\proj{\trace(\exec)}{\actions_\updatealg} \in \lang(\updatealg)$. We say that $\updatealg$ guarantees update consistency for $\mathcal{T}$ if all computations controlled by $\updatealg$ are consistent.

\subsubsection{Impossibility Result}

The blue/green and big flip deployment algorithms share one desirable property: their ability to guarantee update consistency does not depend on semantic details of the underlying system $\mathcal{T}$ nor the specific message values observed during an execution of the system. In particular, they do not make decisions about whether to direct a specific request to an updated worker based on the request's identity or its effect on $\mathcal{T}$'s global state. In other words, they offer universally applicable solutions to update consistency. We call such algorithms \emph{oblivious}.

On the other hand,  blue/green and big flip updates disallow mixed-mode entirely. This begs the following question: do there exist update algorithms that are oblivious, yet allow mixed-mode computations? We show here that such an oblivious update scheme is not possible.

To state our impossibility result precisely, we must define our terms.
We are interested in update \dc{algorithms}\dcout{ implementations} that lead to mixed-mode operation.
Intuitively, a computation $\exec=\gstate_0,\action_0,\dots$ is in mixed-mode if there is interaction between updated and non-updated system components (indirectly via the database). Formally, we say that worker $\worker{} \in \workers$ has been updated before time $k$ in $\exec$ if there exists $i < k$ such that $\action_i \in \uactions_{\worker{}}$. Then $\exec$ is in \textdef{mixed-mode} if there exist $\worker{1},\worker{2} \in \workers$ and $i < j$ such that \kn{worker} $\worker{1}$ has been updated before $i$, \kn{worker} $\worker{2}$ has not been updated before $j$, \kn{and actions} $\action_i = \snd{\worker{i}}{\database}{\mathit{op}_1}$ and $\action_j = \snd{\worker{2}}{\database}{\mathit{op}_2}$ for some database operations $\mathit{op}_1,\mathit{op}_2 \in \messages$.

We say that traces $\alpha$ and $\beta$ are equal up to message values if $\beta$ can be obtained from $\alpha$ by only changing the message values in the send and receive actions in $\alpha$. An update algorithm $\updatealg$ is \textdef{oblivious} if for any $\alpha,\beta$ over $\actions_\updatealg$ that are equal up to message values, $\alpha \in \lang(\updatealg)$ iff $\beta \in \lang(\updatealg)$.

With these definitions, we can now state our impossibility result.
\dc{Intuitively, an oblivious update algorithm cannot exploit knowledge of the particular system or update, and so it cannot direct requests between updated and non-updated workers in a way that ensures that every client perceives an atomic update. Concurrent service by differently-versioned workers can reveal the system's internal inconsistency. Thus, an oblivious update algorithm cannot guarantee update consistency when the computation operates in mixed-mode.}

\begin{theorem}[Impossibility result]
    For $|\messages| \geq 2$, there does not exist an update algorithm that:
    \begin{enumerate}
    \item admits a mixed-mode controlled computation for some system,
    \item is update consistent for all systems, and
    \item is oblivious.
    \end{enumerate}
\end{theorem}

\begin{proof}
By way of contradiction, assume that $\updatealg$ is an oblivious update algorithm that is update consistent for all systems with controllable actions $\actions_\updatealg$. Assume further that $\exec$ is a mixed-mode computation of some system $\mathcal{T}$ with controllable actions $\actions_\updatealg$ such that $\exec$ is controlled by $\updatealg$.
We construct a system $\mathcal{T}'$ with a computation $\exec'$ that is not update consistent but equal to $\exec$ up to message values. 
\dc{Since $\updatealg$ is oblivious, $\exec'$ must also be controlled by $\updatealg$, but this contradicts the assumption that $\updatealg$ is update consistent for all systems.}

For clarity of explanation, in this construction we do not reason explicitly about internal actions occurring in $\exec$ that are not worker update actions. These actions are simply treated as implicit skip actions of every state of the constructed local LTS comprising $\mathcal{T}'$.

As $|\messages| \geq 2$, we let $0$ and $1$ represent two distinct elements of $\messages$. We then define $\mathcal{T}'$ as follows. Each client $\client{}$ consists of a loop that sends a $0$ request to some worker and waits for a $0$ or $1$ response:
  \[ T_{\client{}} = \mu \mathsf{C}.\; \sum_{\worker{} \in \workers} \ccssnd{\worker{}}{0}.\, \sum_{\msg \in \{0,1\}} \ccsrcv{\worker{}}{\msg}.\, \mathsf{C} \enspace.\]
  Every worker $\worker{}$ consists of a loop that either updates the worker or processes a $0$ request from some client. When processing a $0$ request, if $\worker{}$ has not yet updated, it sends operation $0$ to the database and otherwise it sends $1$. In either case, it waits for a $0$ or $1$ response from the database, which it echos to the client:
  \[\arraycolsep=2pt
    T_{\worker{}} =  \begin{array}[t]{rl}
  \mu \mathsf{W}. &
  \sum_{\uaction \in \uactions_{\worker{}}} \uaction.\, (\mu \mathsf{W'}.  \sum_{\client{} \in \clients} \ccsrcv{\client{}}{0}.\, \ccssnd{\database{}}{1}. \, \sum_{\msg \in \{0,1\}} \ccsrcv{\database{}}{\msg}.\, \ccssnd{\client{}}{\msg}.\, \mathsf{W}')\\ 
  + & \sum_{\client{} \in \clients} \ccsrcv{\client{}}{0}.\, \ccssnd{\database{}}{0}. \, \sum_{\msg \in \{0,1\}} \ccsrcv{\database{}}{\msg}.\, \ccssnd{\client{}}{\msg}.\, \mathsf{W}\enspace.
  \end{array}\]
The database accepts $0$ and $1$ operations from any worker. If it receives $0$ after having previously received $1$ from another worker, then it responds with $1$, otherwise with $0$ \dc{(the database can be made passive by shifting activity to the worker)}:
\[\arraycolsep=2pt
  T_\database = \mu \mathsf{DB}.\; 
  \begin{array}[t]{rl}
\sum_{\worker{} \in \workers} &
  \ccsrcv{\worker{}}{0}. \, \ccssnd{\worker{}}{0}.\, \mathsf{DB} \\
  + & \ccsrcv{\worker{}}{1}. \, \ccssnd{\worker{}}{0}.
    \begin{array}[t]{rl}
      (\mu \mathsf{DB}'. & \ccsrcv{\worker{}}{0}.\, \ccssnd{\worker{}}{1}.\, \mathsf{DB}' + \ccsrcv{\worker{}}{1}.\, \ccssnd{\worker{}}{0}.\, \mathsf{DB}')\enspace.
    \end{array}
  \end{array}
\]
Observe that all computations of $\mathcal{T}'$ with an atomic update only contain client responses of the form $\rcv{\client{}}{\worker{}}{0}$. Hence, the same is true for any computation of $\mathcal{T}'$ with a consistent update.
On the other hand, any mixed-mode computation contains at least one client response $\rcv{\client{}}{\worker{}}{1}$. Therefore, $\mathcal{T}'$ does not have any mixed-mode computation that is update consistent.

\dc{Recall that $\exec$ is a mixed-mode computation controlled by $\updatealg$
.}
We \dc{now} map $\trace(\exec)=\action_0,\action_1,\dots$ inductively onto a trace $\alpha=b_0,b_1,\dots$ of $\mathcal{T}'$\dc{, a system which by definition cannot admit an update consistent mixed-mode computation}. Intuitively, the mapping replaces the message value occurring in \dc{each} $a_k$ with $0,1$ to \dc{obtain $b_k$ in a way that produces a valid execution of $\mathcal{T}'$,} leaving all internal actions unchanged. 
\dc{Validity} of $b_k$ is guaranteed using auxiliary functions defined over the already constructed prefix $b_0,\dots,b_{k-1}$ to determine what state $\mathcal{T}'$ will have to be in when executing $b_k$. For example, when $a_k$ is a receive action  $a_k = \rcv{\client{}}{\worker{}}{\msg'}$ for some $\msg'$, the mapping has to inspect the constructed prefix trace to see what value $\msg \in \{0,1\}$ was most recently sent from $\worker{}$ to $\client{}$ and substitute this for $\msg'$ in $a_k$ to obtain $b_k$. Similarly, when $a_k$ is a database response to a worker $a_k = \snd{\database}{\worker{}}{\msg'}$, it has to inspect the prefix to determine what $\worker{}$-operation $\database$ is responding to and whether $\database$ has already seen an operation request from an updated worker.

Formally, the mapping is defined as follows:
\[
  b_k = \begin{cases}
    \snd{\client{}}{\worker{}}{0} & \text{if } \action_k = \snd{\client{}}{\worker{}}{\msg}\\
    \rcv{\client{}}{\worker{}}{\msg} & \text{if } \action_k = \rcv{\client{}}{\worker{}}{\msg'} \land \mathit{lastsnd}(k,\worker{},\client{})=\msg\\
    \snd{\worker{}}{\database}{\msg} & \text{if } \action_k = \snd{\worker{}}{\database}{\msg'} \land \mathit{updated}(k,\worker{})=\msg\\
    \rcv{\worker{}}{\database}{\msg} & \text{if } \action_k = \rcv{\worker{}}{\database}{\msg'} \land \mathit{lastsnd}(k,\database,\worker{})=\msg\\
    \snd{\database{}}{\worker{}}{1} & \text{if } \action_k = \snd{\database{}}{\worker{}}{\msg'} \land \mathit{lastsnd}(k,\worker{},\database)=0 \land \exists \worker{}' \in \workers.\;\mathit{lastrcv}(k,\database,\worker{}')=1\\
    \snd{\database{}}{\worker{}}{0} & \text{if } \action_k = \snd{\database{}}{\worker{}}{\msg'} \land (\mathit{lastsnd}(k,\worker{},\database)=1 \lor \forall \worker{}' \in \workers.\;\mathit{lastrcv}(k,\database,\worker{}')=0)\\
    \rcv{\database}{\worker{}}{\msg} & \text{if } \action_k = \rcv{\database}{\worker{}}{\msg'} \land \mathit{lastsnd}(k,\worker{},\database)=\msg\\
    \action_k & \text{otherwise}
  \end{cases}
\]
where $\mathit{lastsnd}(k,\procA,\procB)$ is $\msg$ if the last send action from $\procA$ to $\procB$ in $b_0,\dots,b_{k-1}$ is $\snd{\procA}{\procB}{\msg}$ and $0$ if no such send actions exist. The function $\mathit{lastrcv}(k,\procA,\procB)$ is defined correspondingly. Similarly, $\mathit{updated}(k,\worker{})$ is $1$ if some $\uaction \in \uactions_{\worker{}}$ occurs in $b_0,\dots,b_{k-1}$, and $0$ otherwise.

By induction, one may construct a computation $\exec'$ of $\mathcal{T}'$ with $\trace(\exec') = \alpha$. By construction, $\exec'$ is equal to $\exec$ up to message values. 
\dc{Since $\exec$ is a mixed-mode computation,} it follows that $\exec'$ is in mixed-mode, 
\dc{and since it is a computation of $\mathcal{T}'$ it cannot be update consistent.} However, because $\updatealg$ is oblivious, $\exec'$ is controlled by $\updatealg$ and, hence, $\updatealg$ is not update consistent for $\mathcal{T}'$. Contradiction.\qedhere
\end{proof}

\section{Commutativity-Based Update Algorithms}
\label{sec:comm-update-algorithms}

In this and the following sections, we give algorithms that guarantee update consistency for mixed-mode updates by exploiting semantic properties of system actions. At a system level, the actions of one client are not directly observed by another client; moreover, from a client's perspective, the service is a black box. This, together with the client-focused view of consistency, opens up possibilities for rearranging or even rewriting service internal actions in a manner that is oblivious to a client, but ensures that she sees results that appear to arise from an atomic update.

\subsection{Orderedness}
It is essential that any rewrite preserve a given client's observations.  
Thus, the requests made by a client cannot be serviced in an observable manner by an alternation of updated and non-updated workers, since there will then not exist a client-equivalent computation for that client with an atomic update. \dcout{This gives us a bare minimum property for ensuring consistency, which} We call \dc{this property} an \textdef{ordered update}: once a client request is serviced by an updated worker, subsequent requests by that client are serviced only by updated workers. \dc{One can think of an ordered update as fixing the atomic update point for each client at the moment they receive their first updated response. Since a client observes the update at that moment, orderedness ensures that they will continue to directly observe service by updated workers. More work is required, however, to ensure that their indirect observations via the database also remain consistent with an atomic update from that point onward.}
\dc{\Cref{alg:ordered} only enforces orderedness but not update consistency. (In fact, it is oblivious.)} 
\dc{\cref{alg:basic-comm} and \cref{alg:comm-update-inv} extend \cref{alg:ordered}, exploiting commutativity and controlling service to ensure that observations via the database are consistent with an atomic update for each client.} 

\dc{Orderedness is necessary when the version of a worker is observable.
Properties like backward- and forward-compatibility obscure the version of a servicing worker, allowing us to manipulate \emph{when} each client observes an update. This renders orderedness unnecessary and even undesirable, since orderedness \emph{fixes} each client's observed atomic update point with their first updated response. 
\cref{alg:backward-compatible} and \cref{alg:forward-compatible} exploit backward- and forward-compatibility to shift each client's observed update point later or earlier, respectively, giving update-consistency without orderedness.}

\dcout{In the previous section, we defined an update algorithm in abstract terms, as a transition system that controls the scheduling of system actions. In this section, we give the algorithms more structure.}
\dc{For our algorithms,} we suppose that every system component -- clients, workers, and the database -- has an associated ``update shell'' \dc{as described in \cref{subsubsec:update-algs}}. This is an auxiliary process that controls the flow of messages in and out of the component,  determines the point at which the component is updated, and tracks the update status of the component. \dc{Although only workers update, client shells can record clients as ``updated'' after receiving their first updated response to enforce orderedness. Likewise,} we assume an ``update manager'' component that controls the flow of messages from clients to workers and instructs workers to update (via the worker shell). The individual shells and the update manager are activated only for the duration of an update.
\dcout{Algorithm~\ref{alg:ordered} enforces orderedness. The algorithms proposed in this and following sections are extensions/refinements of Algorithm~\ref{alg:ordered}. They} 

\knout{Our algorithms add more detail on how the update manager controls the flow of client requests to updated or non-updated workers to ensure update consistency.\cref{alg:basic-comm} accomplishes this by exploiting commutativity of service actions; \cref{alg:backward-compatible} and \cref{alg:forward-compatible} by exploiting backward- and forward-compatibility, respectively; \cref{alg:comm-update-inv} combines these approaches.} 

Each\knout{ of these algorithms} \kn{algorithm} requires an analysis of relevant semantic properties of service actions. Since the set of service actions is static, this analysis is meant to be performed in advance of an update and consulted during runtime.

\begin{algorithm}[t]
\caption{Orderedness\dcout{algorithm}}
\label{alg:ordered}\begin{enumerate}
    \item The update manager sends an ``update'' instruction to a worker.
    \item When the worker shell receives the update instruction, it waits for the worker to complete servicing its current request (if any), then updates the worker process, and sends an ``update complete'' message to the update manager.
    \item Upon processing a client request, an updated worker tags the requesting client as ``updated'' by attaching an update tag to the response message.
    \item On receiving an update-tagged response, a client shell\twcomment{This is the first time that ``client shell'' appears. Perhaps discuss this together with update shells.}\dccomment{We include that clients have an update shell in the discussion, but I added an explicit note that we give clients an update status even though clients themselves do not update.} changes its status to ``updated'' and tags all outgoing requests as arising from an updated client.
    \item \label{algstep:ordered} The update manager directs all requests from updated clients to updated workers. \twout{The update manager may have additional behavior, which is refined in the proposed algorithms.}
    \item The update manager continues to send ``update'' instructions until all workers are updated.
\end{enumerate}
\end{algorithm}

\subsection{Commutativity}
\label{subsec:commutativity}

The first semantic property we  exploit to guarantee update consistency is commutativity.
Intuitively, if two client requests commute, a client cannot identify the order in which the requests were \dc{serviced}\dcout{ performed}. In a case where an updated action was performed before a non-updated action, commutativity can allow an observing client to assume that the actions were performed in the opposite order, hiding mixed-mode behavior from the client. 

Consider our \dc{message} translation update once more. Suppose Alice sends a \dc{message} to Bob via an updated worker, so her \dc{message} is automatically translated. Before checking his inbox, Bob sends a \dc{message} to Charles via a non-updated worker, so no translation is attached. Then, Bob checks his inbox and sees the automatically translated \dc{message} from Alice. Although in the actual computation, Alice's updated send precedes Bob's non-updated send, the two actions are commutative: without a check-inbox action by Bob in between, there is no dependency ordering between the two \dc{message} sends. Thus, Bob may assume that his non-updated \dc{message} send preceded Alice's updated \dc{message} send, giving him the perception of an atomic update.

\subsubsection{Formal Definitions of Commutativity}
We say that action $a$ \textdef{right-commutes} with action $b$ 
if for all initial computation fragments $\alpha;a;\gstate_0;b;\gstate_1$, there exists a state $\gstate_0'$ such that $\alpha;b;\gstate_0';a;\gstate_1$ is also a computation fragment.
\textdef{Left-commuting} is defined similarly. Two actions \textdef{commute} if they both left- and right-commute. We generalize these definitions to commutativity of sequences of actions $\alpha$ and $\beta$ in the expected way.

\subsubsection{Commutativity Algorithm}
\cref{alg:basic-comm} exploits commutativity in a basic manner to ensure update consistency. The algorithm itself is simple and the proof of update consistency for this algorithm seems intuitive, but in fact the formal proof is quite involved. After presenting the algorithm and its intuitive proof, we will build up the technical machinery required to prove update consistency, for this algorithm and for our subsequent algorithms.

Note that by the architectural assumptions on systems, every relay contains a computation fragment whose trace $\delta_r$ is a sequence of database actions of the shape $\rcv{\database}{\worker{}}{\mathit{op}}.(\iactions_\database)^*.\snd{\database}{\worker{}}{\mathit{res}}$. We refer to this trace as the database transaction of the relay (respectively, its associated client request).

\begin{algorithm}[t]
\caption{Basic Commutativity\dcout{algorithm} (Extension of Algorithm \ref{alg:ordered})}
\label{alg:basic-comm}
\begin{enumerate}
    \item \label{algstep:tell-db-updating} At its start, the update manager notifies the database that the update has begun.
    \item \label{algstep:enforce-req-order} During the update, the update manager tags all incoming client requests with a \kn{sequence} number. The database enforces that requests are completed in the assigned order.
    \item The update manager operates as follows:
      \begin{enumerate}
        \item On receiving a request $r$ from an updated client, the manager sends the request to an updated worker, as defined in Algorithm~\ref{alg:ordered}. 
        \item \label{algstep:left-comm}  Upon receiving a request $r$ from a non-updated client, the update manager checks if the database transactions of the current-version of $r$ commute to the left of all prior new-version database transactions. If that is the case, the update manager may assign the request $r$ to any worker process, regardless of worker version. Otherwise, the update manager must assign this request to a new version worker.
    \end{enumerate}
  \end{enumerate}
\end{algorithm}

\paragraph{\dc{Proof Sketch}}
\dc{Intuitively, the commutativity property enforced by \cref{alg:basic-comm}} allows a mixed-mode execution to be sorted after the update is complete, by commuting all non-updated actions to the left (while preserving the ordering between non-updated actions). By commutativity, this (conceptual) sorting guarantees that the end state of the system after sorting is identical to the end state reached by the original update computation.  Thus, to every client, the actual system behavior is indistinguishable from that observed during an atomic update. 

The proof sketch for the algorithm laid out above is fairly straightforward. However, it is not very precise.  A formal proof must justify that every client can perceive a plausible, observationally equivalent computation with an atomic update. For that, we need to define exactly how to ``legally'' rewrite (informally, to sort) a computation, possibly over multiple rewriting steps, in a manner that preserves client observations. Specifically, in the case of commutativity, this requires us to argue that swapping the order of commuting transactions does not introduce a paradoxical time (formally, dependency) loop, and instead produces a valid computation. In the next section, we develop foundational theory to construct such consistency proofs. We then give a formal proof of the update consistency of Algorithm \ref{alg:basic-comm} and present additional algorithms for consistent updates.

\paragraph{\dc{Overhead}}
\cref{alg:basic-comm} enforces that in an update computation, every non-updated action left-commutes with all preceding updated actions. 
\dc{To achieve this property, we must perform a commutativity analysis of available operations in advance of the update; during the update we must impose an order on all database operations (Step \ref{algstep:enforce-req-order}), and consult the results of the commutativity analysis (Step \ref{algstep:left-comm}). 
Commutativity analysis can be complex. Several works~\cite{DBLP:conf/vmcai/PincusK25, veracity, auto-comm-cond} offer automated techniques for generating commutativity conditions. 
Running this analysis in advance prevents serious delays during runtime. 
However, imposing an order on the database can cause serious delays.
If the update manager is distributed, assigning an order also requires a consensus protocol.

Ordering database actions is necessary to ensure that all non-updated actions commute with \emph{prior} updated actions, which guarantees update consistency.
One option to achieve this property while mitigating delays is for the database to accept any request with a number higher than the previously completed request. Requests with a number lower than the previously completed request are rejected by the database, and sent back to the manager for a new number, re-evaluation of commutativity against all requests with lower numbers, and re-assignment to a worker. (This results in stronger commutativity requirements than necessary, since some of those lower-numbered requests will also be rejected by the database.)
Another option is to limit the available operations during the update to ones that are mutually commutative. Then all requests can be sent to any worker, and no database ordering is necessary. This restricts operation during the update, but increases efficiency.
}

\newcommand{\BD}{\mathit{bd}}
\newcommand{\RD}{\mathit{rd}}

\section{Proving Update Consistency}
\label{sec:proving-update-consistency}

In this section, we develop proof principles for showing the consistency of updates. The definition of update consistency requires that for every client $c$, the actual computation is $c$-equivalent to a computation with an atomic update. \tw{To show update consistency for a given client,} we begin with the original computation and apply a sequence of \emph{client-specific rewrites} that constructively transform the original computation to one with an atomic update. \twout{To show termination}\tw{For this argument to be well-founded}, each rewrite must reduce a \twout{well-founded} progress measure. We recast consistency in terms of a partially ordered view of a computation that simplifies proofs by focusing only on the required dependencies between actions and define an appropriate progress measure. This leads to an induction principle for consistency proofs.

\dc{We fix a communicating transition system $\mathcal{T}$ and an update algorithm $\updatealg$; ``computations'' refer to computations of $\mathcal{T}$ controlled by $\updatealg$.}

\subsection{A Partial Order Model} 
\label{subsec:partial-order}
Up until now, we have used an interleaving model for computations, where computations alternate between global states and single actions. However, interleaving is unnecessarily strong as it imposes a total ordering on actions even if they are semantically independent of each other. 
We now introduce a partial order model that retains only the necessary dependencies between actions to simplify our theorems and proofs. This model is adapted from \cite{lamport-time-clocks} and extensions such as \cite{chandy-lamport, mattern1989virtual}.

We define the set of ordered actions $\ordactions_\procA$ of a process $\procA$ by identifying every occurrence of an action in a computation $\exec$ of $T_\procA$ with the trace of the initial computation fragment of $\exec$ ending with that occurrence. Formally, $\ordactions_\procA = \pset{\alpha[0,k]}{k \in \nat \land \alpha \in \lang(T_\procA)}$. We use the same symbols to denote ordered actions as we use for system actions whenever clear from context.
We also often refer to an ordered action $\alpha[0,k]$ as an action and identify it with the system action $\alpha[k]$.
For ordered actions $\action$ and $\actionB$ of $\procA$, we write $\action <_\procA \actionB$ if $\action$ is a prefix of $\actionB$ (informally, $\action$ occurs prior to $\actionB$ in the timeline for process $\procA$). Let $\ordactions_\system = \bigcup_{\procA \in \procs} \ordactions_\procA$.

A \textdef{po-computation (fragment)} is a tuple $\poexec=(\actions, \startstate, \porder)$ where $\actions \subseteq \ordactions_{\mathcal{T}}$ is a set of ordered actions, $\startstate$ is the start state of the po-computation, and $\porder$ gives a partial order over $\actions$ such that every total ordering that satisfies the partial order yields a computation fragment starting in $\startstate$ (when one fills in resulting intermediate states). 
In other words, a po-computation gives ordering constraints on actions necessary to yield a computation fragment starting in $\startstate$. We refer to these computation fragments as the linearizations of $\poexec$.
A well-formed po-computation must have at least one linearization. From this point on, all po-computations discussed are assumed to be well-formed. (The consistency proofs apply to po-computations that are induced by computation fragments, which thus have non-empty sets of linearizations and are therefore well-founded.)

We are interested in po-computations whose partial order is defined by the \textdef{happens-before} relation \cite{lamport-time-clocks} $\porder_{hb}$, where $\action <_{hb} \actionB$ if: 
\begin{itemize}
\item $\action$ is a send action and $\actionB$ is its corresponding receive action (``message-passing order''), or
\item $\action$ and $\actionB$ are both performed by the same process $\procA$, and $\action <_\procA \actionB$ (``program order''), or
\item $\action <_{hb} \actionB$ via the transitive closure of message-passing order and program order.
\end{itemize}
This relation captures the causal ordering of actions in the system.
From this point onward, we assume that all po-computations use the happens-before relation as the partial order, and denote this relation simply as $<$. We lift the happens-before order to sequences of actions, $\alpha < \beta$, to mean that $\alpha$ and $\beta$ are totally ordered by happens-before and all actions in $\alpha$ happen before actions in $\beta$.
Note that if the set $\actions$ in a po-computation is finite, then all linearizations under the happens-before order have the same last state. 

Every computation fragment $\exec$ induces a unique po-computation $\poexec$ via the happens-before relation such that $\exec$ is one of $\poexec$'s linarizations, but all causally independent actions in $\exec$ are left unordered in $\poexec$. The following lemma summarizes a useful property of po-computations. (Proofs omitted from this section can be found in \appendixRef{appendix:formal-model,appendix:proving-update-consistency}. \twcomment{We can cite the arXiv tech report here.})

\begin{lemma}
\label{lem:msg-pass-bet-procs}
    Suppose $\action$ and $\actionB$ are two actions by distinct processes $\procA$ and $\procB$, with $\action < \actionB$. Then, there must exist some actions $\action'$ and $\actionB'$ such that $\action'$ is a send, $\actionB'$ is a receive, and $\action \leq_\procA \action' < \actionB' \leq_\procB \actionB$.
    In other words, there must be (possibly indirect) message passing between $\procA$ and $\procB$.
\end{lemma}

An \textdef{update for a po-computation} $\poexec$ is a set of actions containing all the update actions in $\poexec$.

An \textdef{atomic update for a po-computation} is an update where no update actions are related to each other in the partial order.
While this definition may seem surprising, it is interchangeable with the atomic update definition for computations.
\begin{lemma}
  \label{lem:po-atomic-update}
    A po-computation has an atomic update iff one of its linearizations has one.
\end{lemma}
\begin{proof}
    Only if: Suppose we have no update actions related to each other in the partial order of a po-computation. Then, there exists a total ordering consistent with the partial order such that all update actions occur in one uninterrupted block.
    
    If: Suppose a computation has an atomic update. By definition, all update actions occur in one contiguous block. Consider update actions $a$ and $b$. We show by contradiction that they must be unrelated \dc{in the lifted po-computation}. Assume that $a < b$.  Update actions are not send/receive actions, nor can we have two updates by one process. By Lemma \ref{lem:msg-pass-bet-procs}, we must then have actions $a'$, $b'$ such that $a < a' < b' < b$ where $a'$ is a send and $b'$ a receive. But that contradicts the assumption that $a$ and $b$ appear in an update-only block \dc{in the computation}. Hence, all pairs of updates must be unrelated \dc{in the lifted po-computation}.
\end{proof}

We  lift $\procA$-equivalence and update consistency to po-computations in the expected way. 

\begin{theorem}
\label{thm:full-rewrite}
    Suppose we have a computation $\exec$ with an update $\update$ and $\poexec_u$ is the po-computation induced by $\update$. If for every client $\client{}$ there exists $\poexec'_u$ that is $\client{}$-equivalent to $\poexec_u$ and has an atomic update, then $\exec$ is update-consistent.
\end{theorem}
\begin{proof}
  By definition, $\exec$ is update-consistent if, for every client $\client{}$, there exists a computation with an atomic update that is $\client{}$-equivalent to $\exec$.
  Let $\client{}$ be a client and $\poexec'_u$ a po-computation that is $\client{}$-equivalent to $\poexec_u$ and has an atomic update.
  By assumption and \cref{lem:po-atomic-update}, there exists a linearization $\update'$ of $\poexec_u'$ that is an atomic update. 
  Replacing $\update$ with $\update'$ in $\exec$ (possible as both end in the same global state by equivalence) we obtain a computation with an atomic update that is $\client{}$-equivalent to $\exec$.
\end{proof}

\subsection{Cuts}
\label{subsec:cuts}
While partial order models describe a microservice in terms of causality rather than time, we can reference a particular moment in time in a po-computation $\phi=(\actions,\startstate)$ by choosing a point on the (totally ordered) timeline of each process. A \textdef{cut} \cite{consistent-global-states} is a set of actions $\cut \subseteq \actions$ that includes, for each process, all actions by that process up until a certain point.
$\cut$ is \textdef{consistent} if it is downwards-closed according to the happens-before relation. Consistent cuts in $\poexec$ correspond to reachable states of a corresponding interleaved execution. 

\begin{figure}
    \centering
    \includegraphics[width=0.8\linewidth]{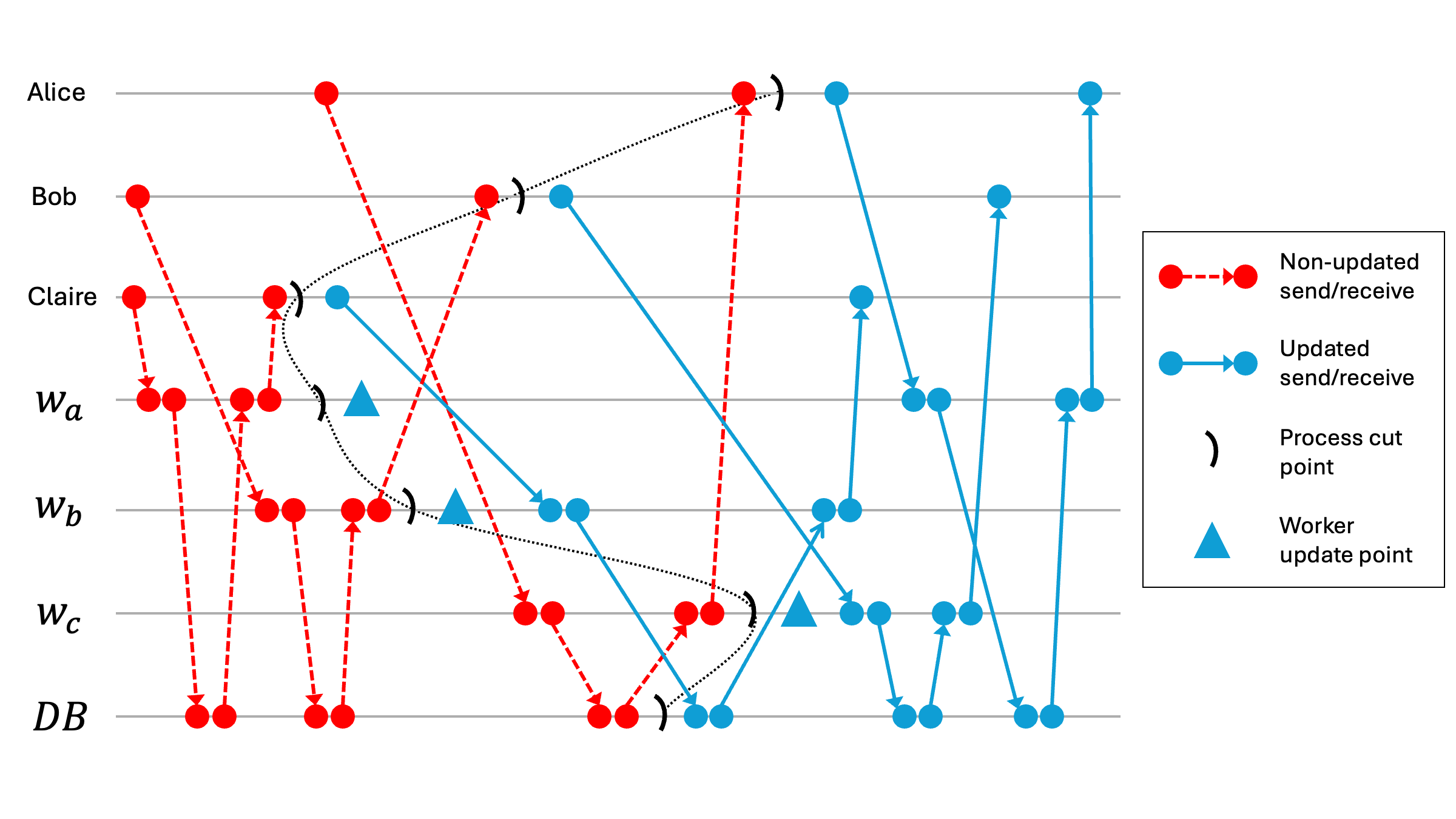}
    \Description{Illustrating an update consistent cut}
    \caption{Space-time diagram showing a global update cut (\cref{sec:progress-measures}), the downwards closure of clients' non-updated receives. Each process's timeline orders actions from left to right. Alice, Bob, and Claire are clients, $w_{a,b,c}$ are workers, and $DB$ is the database. The dotted curve connecting the cut points gives the full cut. The cut is update-consistent: all non-updated actions lie within the cut, while all updated actions lie outside the cut.}
    \label{fig:thm:update-consistent-cut}
\end{figure}

Consistent cuts are a useful tool for proving that a po-computation has an atomic update.
Although only workers undergo update, we call an action ``updated'' if it is part of a relay with an updated worker, and ``non-updated'' \tw{otherwise}\twout{if it is part of a relay with a non-updated worker}.
We define an \textdef{update-consistent cut} as a consistent cut that includes all the non-updated actions and none of the updated actions of every process. (See \cref{fig:thm:update-consistent-cut}.
\dc{For easier visualization, we assign colors to actions: non-updated actions are colored red and updated actions are colored blue}.) 
An update occurs atomically at a certain point in time if all prior actions are non-updated, and all subsequent actions are updated. An update-consistent cut divides actions in this way, representing an atomic update point.

\begin{theorem}
\label{thm:update-consistent-cut}
    A po-computation has an atomic update iff it admits an update-consistent cut. 
\end{theorem}
\begin{proof}
  Only if direction: let $\poexec$ be a po-computation that has an atomic update. Choose as the cut the set of all non-updated actions of $\poexec$. This cut is consistent by the assumption that $\poexec$ has an atomic update.
  
    If direction: 
    Suppose the po-computation does not have an atomic update. Then there exists a pair of update actions, $a$ and $b$, such that $a < b$. By Lemma \ref{lem:msg-pass-bet-procs}, we must then have actions $a'$ and $b'$ where $a < a'$ by program order, $b' < b$ by program order, and $a' < b'$. 
    By assumption, we have an update-consistent cut. Since $a'$ is an updated action, it is not in the update-consistent cut, and since $b'$ is a non-updated action, it is in the update-consistent cut. But since $a' < b'$, the cut is not downwards-closed, contradiction.
\end{proof}

\dcout{To better visualize the constructions, we assign colors to actions: non-updated actions are colored red and updated actions are colored blue (Figure~\ref{fig:thm:update-consistent-cut}). We then have the following properties.}

\subsection{Progress Measures}
\label{sec:progress-measures}

Now that we have rebuilt our formal framework in the partial order model, 
we can use the notion of cuts to give a measure for how ``close'' a computation is to having an atomic update. An intuitive way for measuring this is by observing the point at which all workers are updated, and noting how many actions prior to that point revealed that the update was not atomic. Every action that revealed the non-atomicity of the update brings the update ``farther'' from an atomic one. 

To capture this expected atomic update point, we define a \textdef{global update cut} of a po-computation $\poexec$, which is the downwards closure of the set of all clients' non-updated receive actions in $\poexec$. (See \cref{fig:thm:update-consistent-cut}.)
This collects all of the actions performed by the system processes up until every client receives their last non-updated response, which will include all non-updated actions.

\begin{lemma}
\label{lem:red-inside-global-update-cut}
    \dc{All red actions are inside the global update cut.}
\end{lemma}

For \dc{the global update cut}\dcout{ this point} to be perceived by clients as an atomic update point, the cut should be update-consistent\dcout{: there should be}, \dc{with} only non-updated actions before this point and only updated actions after\dcout{ this point}.

\begin{definition}
    The \textdef{cut progress measure} of a po-computation is the number of updated actions inside its global update cut.
\end{definition}

\begin{theorem}
\label{thm:cut-0-atomic}
    \dcout{Suppose we have a po-computation $\poexec$ with an update. }If the cut progress measure of \dc{a po-computation}\dcout{ $\poexec$} is 0, then it has an atomic update.
\end{theorem}

The cut progress measure by itself is insufficient. Two cuts may both have cut progress measure $1$, showing that each includes exactly one updated action. But in one cut, the updated action may be closer to the cut point on its process timeline than the other updated action is to its timeline cutpoint. That is, the first cut is closer to being ``sorted'' than the second. The second progress measure addresses such situations,\dcout{ but measures} \dc{by measuring} sortedness on the timeline of \emph{database} actions. 
The database is the single point of interaction between distinct workers and clients. Thus, the database exposes the system's mixed-mode operation, and suffices for the sortedness measure.

\begin{definition}
    For each non-updated \emph{database action} $d$ in a po-computation $\poexec$, let $n_d$ be the number of updated database actions that precede $d$. The \textdef{sort progress measure} of a po-computation is the sum of all $n_d$. (In other words, the sort measure is the number of \emph{inversions}, where non-updated actions should precede updated actions.)
\end{definition}

\dc{Proving that the sort progress measure suffices for update-consistency requires this lemma:}
\begin{lemma}
\label{lem:b-rr-distinct-workers}
    \dc{Suppose blue action $b$ is inside of the global update cut. Let $\mathit{rr}$ denote a red client receive action such that $b$ is in the downwards closure of $\mathit{rr}$. Then $b$ and $\mathit{rr}$ are serviced by distinct workers.}
\end{lemma}

\begin{theorem}
\label{thm:sort-0-consistent-cut}
    \dcout{Suppose we have a po-computation $\poexec$ with an ordered update. }If the sort progress measure of \dc{a po-computation}\dcout{ $\poexec$} is 0, then it\dcout{ admits an update-consistent cut} \dc{has an atomic update}.
\end{theorem}

\dc{\cref{thm:sort-0-consistent-cut} relies on our architectural assumptions that clients are sequential, and that every relay contains one database access (modeling no accesses with a ``skip'' access). These assumptions allow us to use the database as a proxy for the full microservice, since it represents the actions of each client, and covers every relay.}

Both the sort and the cut progress measures capture update atomicity at measure 0, and so we can rely on either of them to show update atomicity after repeated decrements. We combine the two measures lexicographically, with the cut measure having higher priority. 

\begin{definition}
The \textdef{progress measure} of a po-computation $\poexec$, denoted $\pmeasure(\poexec)$ is the lexicographically ordered pair of the cut progress measure and sort progress measure of $\poexec$.
\end{definition}

\subsubsection{Induction Principle}
We are now equipped with a progress measure to capture how close a computation is to having an atomic update. Because the ordering on the progress measure is well-founded, this gives us an induction principle for proving update consistency.

\begin{definition}
    \tw{
    A \textdef{client rewrite strategy} $(\rewriteinv,\rewrite)$ consists of a \emph{rewrite invariant} $\rewriteinv$ and a po-computation transformation $\rewrite$, parametric in a client $\client{}$ such that: (1) $\rewriteinv$ is a set of po-computations that includes those controlled by $\updatealg$, and (2) for every client $\client{}$ and any po-computation $\poexec \in \rewriteinv$ where neither component of $\pmeasure(\poexec)$ is $0$, $\rewrite_{\client{}}(\poexec)$ produces a po-computation $\poexec' \in \rewriteinv$ that is $\client{}$-equivalent to $\poexec$, ends in the same global state, and reduces the progress measure, i.e., $\pmeasure(\poexec') < \pmeasure(\poexec)$.} 
\end{definition}

\begin{theorem}
  \label{thm:c-rewrite-strategy-update-consistent}
  \tw{If there exists a client rewrite strategy, then $\updatealg$ is update-consistent for $\mathcal{T}$.}
\twout{Suppose we have an update computation $\exec$ with an induced po-computation $\poexec$. If there is a $\client{}$-rewrite strategy for every client $\client{}$ then $\exec$ is update-consistent.}
\end{theorem}

\begin{proof}
  \tw{Let $(\rewriteinv,\rewrite)$ be a client rewrite strategy and $\client{}$ be a client.}
  We show by induction on the progress measure that \tw{every $\poexec \in \rewriteinv$} is \dc{update-consistent for $\client{}$}\dcout{ $\client{}$-equivalent to some po-computation with an atomic update}.
  
  The base case is where $\pmeasure(\poexec)$ is $0$ in one of its components. Then $\poexec$ has an atomic update by either \cref{thm:cut-0-atomic} or \cref{thm:sort-0-consistent-cut}, and is hence update-consistent. 

  Otherwise, let $\poexec'=\rewrite_{\client{}}(\poexec)$ be the po-computation obtained through the rewrite strategy applied to $\poexec$. We then have $\pmeasure(\poexec') < \pmeasure(\poexec)$ \tw{and $\poexec' \in \rewriteinv$}. By the induction hypothesis, $\poexec'$ \dc{is update-consistent for $\client{}$}\dcout{has an atomic update}. 
  That is, for client $\client{}$, there is a po-computation $\poexec''$ that has an atomic update and is $\client{}$-equivalent to $\poexec'$. As $\poexec'$ is $\client{}$-equivalent to $\poexec{}$, it follows that $\poexec$ and $\poexec''$ are $\client{}$-equivalent.

  As this applies to every client $\client{}$ and $\rewriteinv$ contains all po-computations controlled by $\updatealg$,  algorithm $\updatealg$ is update-consistent by \cref{thm:full-rewrite}.
\end{proof}

\subsection{Commutative Rewrites}
\label{subsec:comm-rewrites}
\dc{We now prove consistency of \cref{alg:basic-comm}. The proof relies on the algorithm's orderedness, which gives the following property.}

\begin{lemma}
\label{lem:b-rr-distinct-clients}
    \dc{Suppose we have an ordered update with some blue action $b$ inside of the global update cut. Let $\mathit{rr}$ denote a red client receive action such that $b$ is in the downwards closure of $\mathit{rr}$. Then $b$ and $\mathit{rr}$ service distinct clients.}
\end{lemma}

\begin{theorem}\label{thm:single-comm-rewrite}
    Algorithm \ref{alg:basic-comm} guarantees a consistent update.
\end{theorem}
\begin{proof}

  We establish this through a rewrite strategy $(\rewriteinv,\rewrite)$ defined as follows. \tw{The rewrite invariant $\rewriteinv$ consists of all po-computations $\poexec$ such that (1) $\poexec$ is ordered (i.e., each client timeline consists of non-updated actions followed by updated actions), and (2) every non-updated database transaction in $\poexec$ is left-commutative with all prior updated database transactions. Consider a computation $\poexec$ controlled by the algorithm. Step~\ref{alg:ordered} of the algorithm ensures property (1). Moreover, by Step~\ref{algstep:enforce-req-order} of the algorithm, the update manager is aware of the total ordering on the database, and so Step~\ref{algstep:left-comm} guarantees property (2). Hence, $\poexec \in \rewriteinv$.}

  \tw{The po-computation transformation $\rewrite$ is then defined as follows.}
  Consider a computation $\poexec \in \rewriteinv$ and a client $\client{}$. If \dcout{the progress measure is not $0$ in either component, there exists a non-updated database transaction $\delta_r$ in $\poexec$ which left-commutes with an immediately preceding updated database transaction $\delta_b$.} 
  \dc{neither component of the progress measure is $0$, then there must be some updated database transaction in $\poexec$ which precedes a non-updated database transaction (contributing towards a non-zero sort measure), and there must be some such adjacent pair. 
  Call this updated transaction $\delta_b$ and this non-updated transaction $\delta_r$. 
  By property (2) of $\rewriteinv$, $\delta_r$ left-commutes with $\delta_b$.}
  The rewrite $\rewrite_{\client{}}(\poexec)$ exchanges the order of the two transactions.

  \knout{By our architectural assumptions on the database process, there does not exist any action $\action$ in $\poexec$ such that $\delta_{r,1} < \action < \delta_{r,2}$ for any non-trivial split $\delta_{r,1}\delta_{r,2} = \delta_r$ (and similarly for $\delta_b$). For the remainder of this proof, we therefore treat $\delta_r$ and $\delta_b$ as if they were atomic (ordered) actions $\mathit{rd}$ and $\mathit{bd}$, respectively. (Technically, the swap replaces each ordered database action $d$ in $\poexec$ that occurs in $\mathit{rd}, \mathit{bd}$ or satisfies $\mathit{bd} < d$, with a modified ordered action $d'$. The modification from $d$ to $d'$ reflects the new sequence of database actions obtained after the swap of $\mathit{rd}$ and $\mathit{bd}$. This is to ensure that we indeed obtain $\mathit{rd} < \mathit{bd}$ according to the program order of the database, and also preserve all happens-before relationships with respect to database actions that succeed $\mathit{rd}$ and $\mathit{bd}$ in $\poexec$. This modification is an order isomorphism with respect to $<_\database$, with the exception that it inverts the order of actions between $\mathit{rd}$ and $\mathit{bd}$. For ease of notation, we continue to refer to the modified ordered actions $d'$ as if they were members of $\poexec$.)}
  \kn{From the formal model of the database process in \cref{subsec:architect}, we can consider the database to act atomically in receiving a request from a worker, acting on it, and sending back a response. We therefore treat $\delta_r$ and $\delta_b$ as if they were atomic (ordered) actions $\mathit{rd}$ and $\mathit{bd}$, respectively. }

  We need to show that this exchange: (I) produces a valid $\client{}$-equivalent po-computation, (II) strictly decreases the progress measure, and (III) preserves the \tw{rewrite invariant $\rewriteinv$}\twout{property that non-updated actions commute to the left over prior updated actions}. The claim then follows from Theorem \ref{thm:c-rewrite-strategy-update-consistent}. We tackle the required conditions in reverse order.

\textbf{III. \tw{Preservation of $\rewriteinv$}\twout{Non-updated actions commute left}.} 
\dc{First, left-commutativity is a property of actions and does not depend on a specific computation.} \tw{Moreover, the exchange only decreases the set of pairs that need to left-commute. Hence, property (2) of $\rewriteinv$ is preserved. Second, the exchange does not modify the client timelines. Hence, property (1) is preserved.}
\dcout{This condition is a global property of the computation, which is unaffected by the interchange of the adjacent actions $\BD$ and $\RD$.} 

\textbf{II. Progress measure decreases.} The reordering clearly strictly decreases the sort measure. Thus, we have to show that the cut measure does not increase. To do so, we show that the cut $C'$ after reordering is a subset of the cut $C$ prior to reordering. Hence, the number of blue (updated) actions in $C'$ is at most the number of updated actions in $C$. 

Consider an action $y$ in $C'$. By the definition of the cut, action $y$ is in the downward closure of the non-updated actions of clients. (Recall that\dcout{ by the orderedness condition,} \dc{this update is ordered, so} a client timeline consists of non-updated actions followed by updated actions.) Thus, there is a chain of dependencies $\gamma = y < d_1 < \ldots d_k < r$ that links $y$ to some non-updated receive action $r$ on some client $c$. We consider several cases. (Recall that a database transaction consists of three actions: receiving the operation from a worker, performing the operation on the database, then returning the result to the sending worker.)

(A) $\gamma$ does not contain events from either $\RD$ or $\BD$. In this situation, the chain $\gamma$ also exists in the original computation; thus, $y$ is in $C$.

(B) $\gamma$ contains only events for $\RD$ but not $\BD$. Then $\gamma$ has the form $\alpha;m;\RD;\beta$ where $m$ is the previous event on the database timeline or the form $\alpha;\RD;\beta$ where the last event of $\alpha$ is on a worker timeline.  In the first case, one can construct $\delta = \alpha;m;\BD;\RD;\beta$; in the second, let $\delta = \gamma$. In each case, the chain $\delta$ exists in the original computation, so $y$ is in $C$.

(C) \dc{$\gamma$ contains events for $\BD$. We can show that $\gamma$ must contain a database action, say $x$, that is beyond $\BD$ in $\gamma$; hence, as $y < \RD$ or $y < \BD$ in $\varphi$, we can form the chain $y < x < r$ in $C$. See \appendixRef{appendix:proving-update-consistency} for details.}

\dcout{$\gamma$ contains events for both $\RD$ and $\BD$. Then $\gamma$ has the form $\alpha;m;\RD;\BD;k;\beta$ where \dc{$m$ is the previous event on the database timeline and} $k$ is the event where the worker receives the response from $\BD$, or the form $\alpha;m;\RD;\BD;n;\beta$ where $n$ is the event on the database timeline after $\BD$. In the first case, we construct $\delta = \alpha;m;\BD;k;\beta$ as a chain in the original computation. In the second case, we construct $\delta = \alpha;m;\BD;\RD;n;\beta$ as a chain in the original computation. In either case, $y$ is in $C$.
}

\textbf{I. Produces valid $\client{}$-equivalent po-computation.} The interchange does not affect the actions of the client $\client{}$, and by commutativity, it leaves the final global state unchanged, so the transformed ordering is $\client{}$-equivalent to the original. \dcout{By assumption, we start with a quiescent update. The rewrite does not interchange the order of actions on a worker, so the transformed order is also quiescent.}

The difficult part of the proof is to establish that the transformed ordering is a valid computation -- i.e., that the transformation does not create a ``time loop''; more formally, a cycle of dependencies. This requires a case analysis.

\dc{We say} \dcout{By} ``directly causally related'' \dc{to}\dcout{, we} mean two actions that are either consecutive actions on the same process, or a send action and its corresponding receive action. All other relations in the partial order are from the transitive closure of those relations, hence ``indirectly'' related.  In this proof, we treat a cycle as composed of a sequence of directly causally related actions.

The proof is by contradiction. Suppose that flipping the partial order to obtain $\mathit{rd} < bd$ does induce a cycle in the order. We show that any such cycle would induce a cycle in the original computation, which is impossible. \dc{We perform case analysis on the possible events in the cycle.}

\textbf{(A) \dc{Neither $rd$ nor $bd$ is in the cycle.}} \dcout{If this cycle does not include $\mathit{rd}$ or $bd$,} The cycle must have been in the order originally, contradicting the assumption that we begin with a proper partial order. \dcout{So the cycle must include at least one of $\mathit{rd}$ or $bd$.}

\textbf{(B) $bd$ is in the cycle.} For a cycle to exist, we must have a sequence $\gamma=q; \alpha; bd; \beta; q$, where $\alpha$ and $\beta$ are sequences of actions and q is an action. 
    
    Action $bd$ is a database action. By \dc{direct causal order} \dcout{the system architecture and the definition of the partial order}, the last action of $\alpha$ must be either the preceding database action, or the send action of the worker (say, $wb$) who sent the operation request for $bd$ to the database. By assumption, the preceding database action is $\mathit{rd}$. 
    Likewise, the first action of $\beta$ must be either the next database action (say, $x$), or the receive action of $wb$ receiving the results sent by the database in action $bd$.
    This gives us a total of four possible cycle shapes for $\gamma$ (where $\alpha'$ is $\alpha$ without its last action and $\beta'$ is $\beta$ without its first action):
    \begin{enumerate}
        \item \label{enum:rd-bd-x} $q; \alpha'; \mathit{rd}; bd; x; \beta'; q$
        \item \label{enum:rd-bd-rcvbd} $q; \alpha'; \mathit{rd}; bd; wb \text{-receive-} bd; \beta'; q$
        \item \label{enum:sendbd-bd-x} $q; \alpha'; wb \text{-send-} bd; bd; x; \beta'; q$
        \item \label{enum:sendbd-bd-rcvbd} $q; \alpha'; wb \text{-send-} bd; bd; wb \text{-receive-} bd; \beta'; q$
    \end{enumerate}
    We proceed by case analysis on the above cycle shapes.
    
    \textbf{Case (\ref{enum:rd-bd-x})} $q; \alpha'; \mathit{rd}; bd; x; \beta'; q$.
    The last action of $\alpha'$ can be either the preceding database action (say, $w$), or the send action of the worker (say, $wr$) who sent the operation request for $\mathit{rd}$ to the database. This gives us two subcases (where $\alpha''$ is $\alpha'$ without the last action).

\begin{figure}[t]
  \begin{minipage}[t]{0.35\textwidth}
    \centering
    \includegraphics[width=\textwidth]{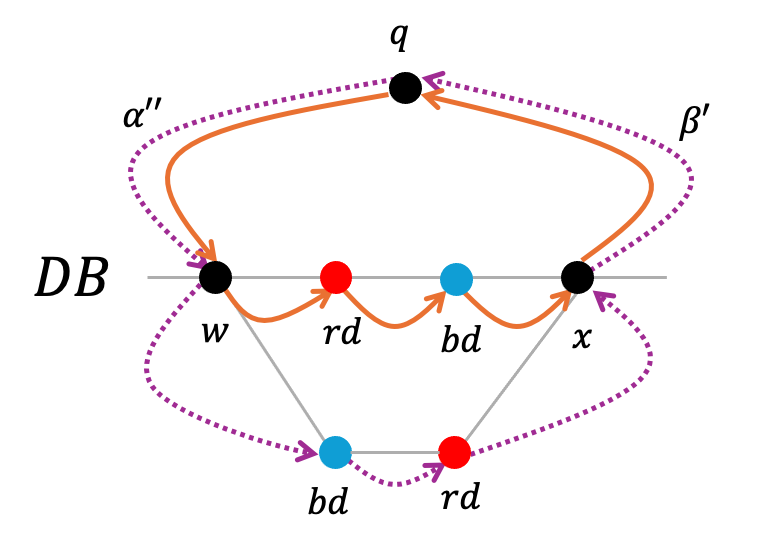}
    \subcaption{Subcase (\ref{enum:rd-bd-x}a)}
    \label{fig:thm:single-comm-rewrite:1a-b-cycle}
  \end{minipage}%
  \begin{minipage}[t]{0.35\textwidth}
  \centering
    \includegraphics[width=\textwidth]{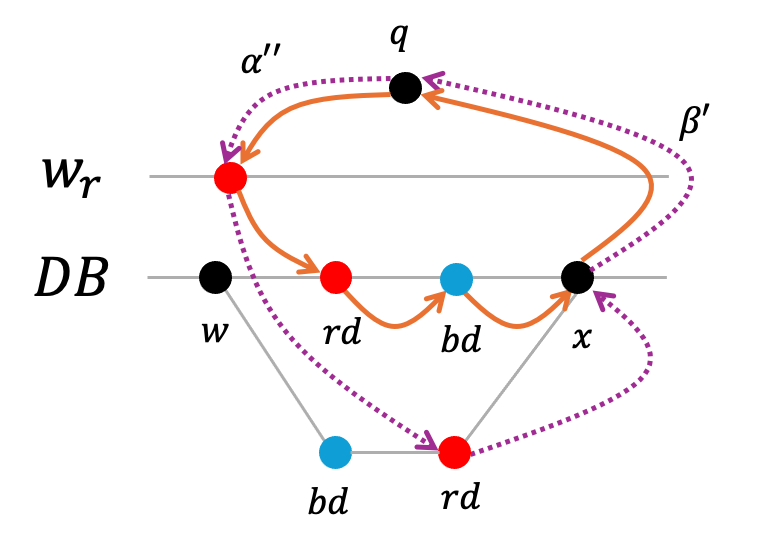}
    \subcaption{Subcase (\ref{enum:rd-bd-x}b)}
    \label{fig:thm:single-comm-rewrite:1b-b-cycle}
  \end{minipage}%
  \begin{minipage}[t]{0.3\textwidth}
    \centering
    \includegraphics[width=\textwidth]{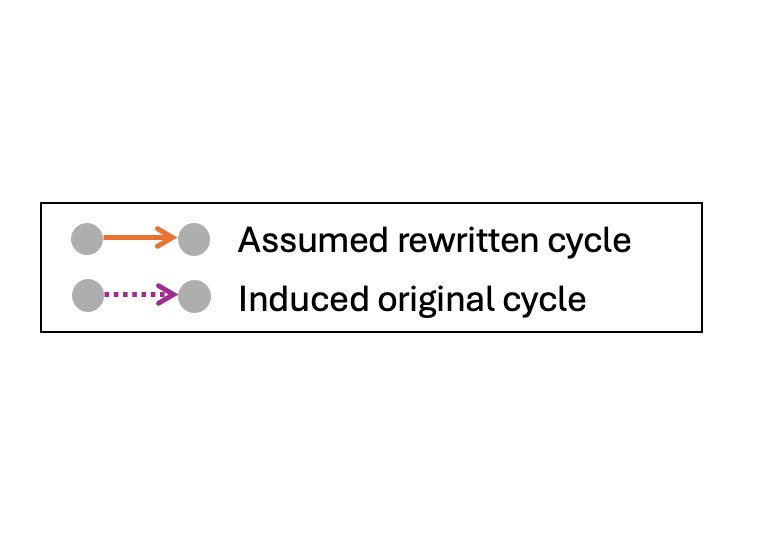}
  \end{minipage}
  \caption{Illustrations of Case (\ref{enum:rd-bd-x}) in the proof of \cref{thm:single-comm-rewrite}, Condition~I, Part~(B). The main database timeline depicts the rewritten computation with $rd < bd$; the original order of $bd < rd$ is depicted below it.}
  \vspace*{-1em}
\end{figure}
    
    \textbf{Subcase (\ref{enum:rd-bd-x}a)} $q; \alpha''; w; \mathit{rd}; bd; x; \beta'; q$.
    See \cref{fig:thm:single-comm-rewrite:1a-b-cycle}.
    Let us consider the subsequence $w; \mathit{rd}; bd; x$. Action $w$ precedes $\mathit{rd}$ and $bd$, and so is unaffected by the order swap. By assumption, the state of the database after $\mathit{rd};bd$ is equal to the state of the database after $bd; \mathit{rd}$ (the original order). Therefore, database action $x$ is identical in either ordering (along with all subsequent actions). The order of $\mathit{rd}$ and $bd$ are swapped in the original, with $w < bd < \mathit{rd} < x$, so the original order must have the cycle $\delta = q; \alpha''; w; bd; \mathit{rd}; x; \beta'; q$.

    \textbf{Subcase (\ref{enum:rd-bd-x}b)} $q; \alpha''; wr \text{-send-} \mathit{rd}; \mathit{rd}; bd; x; \beta'; q$.
    See \cref{fig:thm:single-comm-rewrite:1b-b-cycle}. 
    This subcase is similar to the argument in the previous subcase. The only causal relation that changes in the original ordering is that $bd$ is no longer greater than $wr \text{-send-} \mathit{rd}$, since it is no longer greater than $\mathit{rd}$. This gives us that in the original order we have the cycle $\delta=q; \alpha''; wr \text{-send-} \mathit{rd}; \mathit{rd}; x; \beta'; q$, where $bd$ is removed from the cycle but the cycle is otherwise unchanged.

    \textbf{Case (\ref{enum:rd-bd-rcvbd})} $q; \alpha'; \mathit{rd}; bd; wb \text{-receive-} bd; \beta'; q$.
    We again have two subcases.

    \textbf{Subcase (\ref{enum:rd-bd-rcvbd}a)} $q; \alpha''; w; \mathit{rd}; bd; wb \text{-receive-} bd; \beta'; q$.
    By reasoning similar to that in the first case, we have that the original ordering must have the cycle $\delta=q; \alpha''; w; bd; wb \text{-receive-} bd; \beta'; q$.
    
    \textbf{Subcase (\ref{enum:rd-bd-rcvbd}b)} $q; \alpha''; wr \text{-send-} \mathit{rd}; \mathit{rd}; bd; wb \text{-receive-} bd; \beta'; q$.
    This is the most complex case. 
    \dc{For clarity, we shift the start of the cycle to obtain: 
    $bd; wb \text{-receive-} bd; \beta'; q; \alpha''; wr \text{-send-} \mathit{rd}; \mathit{rd}; bd$.}

    \dcout{Suppose that a cycle does not exist in the original ordering, and so it is induced purely by the swap of $bd; \mathit{rd}$ to $\mathit{rd}; bd$.}   
    In the original ordering, we have $bd < \mathit{rd}$, but the direct causal relations between all other actions \dc{in the cycle} remain the same. 
    \dc{Thus, we have} \dcout{This gives us:} 
    $bd; wb \text{-receive-} bd; \beta'; q; \alpha''; wr \text{-send-} \mathit{rd}; \mathit{rd}$ \dc{in the original computation}\dcout{(shifting the start of the cycle to $bd$)}. But, we do not have $\mathit{rd} < bd$ to create the cycle. \dc{It appears that the cycle induced by the swap does not exist in the original computation.}
    
    \dc{Let us examine the cycle induced by the swap more closely.}
    \dcout{Since} The \dcout{sequence $bd; wb \text{-receive-} bd; \beta'; q; \alpha''; wr \text{-send-} \mathit{rd}; \mathit{rd}$} 
    \dc{actions in the sequence $\beta'; q; \alpha''$ create the relation $wb \text{-receive-} bd < wr \text{-send-} \mathit{rd},$}
    indirectly relat\dc{ing}\dcout{es} $bd < \mathit{rd}$. We \dc{then} obtain a cycle when we swap $bd; \mathit{rd}$ on the database to $\mathit{rd}; bd$: we \dcout{now} have $\mathit{rd} < bd$ from the swap and $bd < \mathit{rd}$ from the indirect ordering.

    What actions could be present in $\beta'; q; \alpha''$ to create the indirect relation $wb \text{-receive-} bd < wr \text{-send-} \mathit{rd}$? 
    By Lemma \ref{lem:b-rr-distinct-workers}, $wb$ and $wr$ are distinct. So, the causal relation cannot arise from $wb \text{-receive-} bd$ and $wr \text{-send-} \mathit{rd}$ occurring on the same process; it must arise through send/receive messages between processes. The system architecture dictates that distinct workers can only be indirectly related through interactions with the same client, and through interactions with the database. 
    By assumption, $bd$ and $\mathit{rd}$ are consecutive database actions. So, $\beta'; q; \alpha''$ cannot contain additional database actions to relate $bd < \mathit{rd}$; then we would have database actions in between $bd$ and $\mathit{rd}$, which contradicts our assumption that they are consecutive. (These database actions cannot be to the left or the right of $\BD;\RD$ as that would form a cycle in the original $\poexec$.)
    
    Therefore, the indirect relation $bd < \mathit{rd}$ must arise from interactions with the same client. \dc{Since this is an ordered update by definition of $\rewriteinv$, we can apply}\dcout{By} Lemma \ref{lem:b-rr-distinct-clients} \dc{to conclude that} $wr$ and $wb$ are servicing distinct clients. So, the interactions that indirectly relate $bd < \mathit{rd}$ must be between these clients, or via a third distinct client. By assumption, all clients are independent, and only interact with other clients indirectly via the database. Then to relate $bd < \mathit{rd}$ via interactions between distinct clients, we would require database actions in between to relate the distinct clients. (These actions cannot be to either side of $\BD;\RD$ as that would create a cycle in the original $\poexec$.) But this again contradicts our assumption that $bd$ and $\mathit{rd}$ are consecutive actions on the database. 

    \textbf{Case (\ref{enum:sendbd-bd-x})} $q; \alpha'; wb \text{-send-} bd; bd; x; \beta'; q$.
    From an analysis similar to the first case, this would induce a cycle $\delta=q; \alpha'; wb \text{-send-} bd; bd; \mathit{rd}; x; \beta'; q$ in the original computation. 
    
    \textbf{Case (\ref{enum:sendbd-bd-rcvbd})} $q; \alpha'; wb \text{-send-} bd; bd; wb \text{-receive-} bd; \beta'; q$.
    This cycle is unaffected by the swap and must exist in the original order.

\textbf{(C) Only $\mathit{rd}$ is in the cycle.}
    Now, suppose $\mathit{rd}$ is part of the cycle induced by the swap. This has the shape $\gamma=q; \alpha; \mathit{rd}; \beta; q$, where $\alpha$ and $\beta$ are sequences of actions and $q$ is an action. There are two cases to consider:
    \begin{enumerate}
    \item \label{enum:x-rd-rcvrd} $q; \alpha'; x; \mathit{rd}; wr \text{-receive-} \mathit{rd}; \beta'; q$. In this case, we have $\delta=q;\alpha';x;\BD;\RD;wr \text{-receive-} \mathit{rd}; \beta'; q$ is a cycle in the original po-computation.
      
    \item \label{enum:sendrd-rd-rcvrd} $q; \alpha'; wr \text{-send-} \mathit{rd}; \mathit{rd}; wr \text{-receive-} \mathit{rd}; \beta'; q$. Here, we have that same cycle is also present in the original po-computation. 
    \end{enumerate}
    
    As any cycle present in the reordered computation induces a cycle in the original computation (which is impossible as the original computation is a partial order), we can conclude that the reordered dependencies cannot result in time loops. 
\end{proof}

\section{Compatibility-Based Update Algorithms}
\label{sec:compat-update-algorithms}
In this section, we will present further algorithms that guarantee update-consistency by ensuring that actions in a computation can be rewritten -- not just reordered -- into client-equivalent atomic update computations for each client. These algorithms rely on the semantic properties of backward- and forward-compatibility, both alone and in combination with commutativity.

\subsection{Compatibility Algorithms}
We present two algorithms for update consistency based on backward- and forward-compatibility, which exploit clients' mutual independence and their ignorance of internal service behavior.

Colloquially, when an update is backward-compatible, old-version requests can be serviced by the updated system in a manner consistent with the original version. 
For the purpose of semantic analysis for update consistency, we will discuss backward-compatibility at the level of individual requests rather than an entire update.
We say that a request is \emph{backward-compatible} if service by an updated worker appears identical to service by a non-updated worker from a client's perspective.

Similarly, forward-compatibility colloquially refers to when new-version requests can be processed by non-updated workers (though this is often limited to coherent error messages). We say that a request is \emph{forward-compatible} if service by a non-updated worker appears identical to service by an updated worker from a client's perspective.

We can achieve compatibility via a \emph{translation layer} by the update manager, if updated actions can be simulated by a sequence of non-updated actions or vice versa. For example, suppose that in our \dc{messaging} service, a user can include a bulleted list in their \dc{message} by manually sending a ``format'' request. We update the service to automatically create a bulleted list format when users type an asterisk on a new line, unless the asterisk is preceded by a backslash. An update manager can enforce forward-compatibility by translating requests with an asterisk on a new line to explicit ``format'' requests, so that requests serviced by non-updated workers appear to be updated. Likewise, it can enforce backward-compatibility by inserting a backslash to prevent the asterisk from being read as a format request, so that requests serviced by updated workers appear to be non-updated.

Both compatibility definitions give a client-oriented view of compatibility, and we will exploit these semantic properties to guarantee update consistency.

\subsubsection{Formal Definitions of Compatibility}

Assume that we are given a \emph{backward-translation} partial function $B: \messages \rightarrow \messages^*$ that maps requests to sequences of requests. 
A request $r$ is \textdef{backward-compatible under translation} if an updated worker servicing the sequence of requests $B(r)$ (atomically) starting in global state $\gstate$ results in the same final global state and client projections as a non-updated worker servicing $r$ in global state $\gstate$.

Analogously, assume $F: \messages \rightarrow \messages^*$ is a \emph{forward-translation} partial function that maps requests to sequences of requests.
A request $r$ is \textdef{forward-compatible under translation} if a non-updated worker servicing the sequence of requests $F(r)$ (atomically) starting in global state $\gstate$ results in the same final global state and client projections as an updated worker servicing $r$ in global state $\gstate$.

\subsubsection{Backward- and Forward-Compatibility Algorithms}
\dc{Unlike \cref{alg:basic-comm}, the compatibility algorithms do not enforce an ordered update, so that the perceived point of atomic update can be manipulated for update consistency.}
\Cref{alg:backward-compatible} takes advantage of backward-compatibility to ensure update consistency by requiring that only backward-compatible actions are sent to updated workers. The update then appears atomic with the last updated worker. 
(The algorithm tests whether a request is backward-compatible as the translation function is partial.)

\Cref{alg:backward-compatible} can be flipped to ensure update consistency by requiring that only forward-compatible actions are sent to non-updated workers (\cref{alg:forward-compatible}), to give the appearance that the system updates atomically with the first updated worker.

\begin{algorithm}[t]
\caption{Backward-Compatibility\dcout{algorithm} \dcout{(extension of Algorithm~\ref{alg:ordered})}}
\label{alg:backward-compatible}
\begin{enumerate}
    \item The update manager operates as follows: \\
    \draftonly{\begin{enumerate}
        \item} \dcout{On receiving a request $r$ from an updated client, the manager sends the request to an updated worker, as defined in Algorithm~\ref{alg:ordered}.} 
        \draftonly{\item} \label{algstep:backward-compatible} On receiving a request $r$ from a non-updated client, the update manager checks if $r$ is backward-compatible under a given translation $B$.
        If it is, then the update manager may assign $B(r)$ (treated atomically) to a new worker process or $r$ to an old worker process.
        Otherwise, $r$ must be assigned to an old version worker process.
    \draftonly{\end{enumerate}}
\end{enumerate}
\end{algorithm}

\begin{algorithm}
\caption{Forward-Compatibility\dcout{algorithm} \dcout{(extension of Algorithm~\ref{alg:ordered})}}
\label{alg:forward-compatible}
\begin{enumerate}
    \item The update manager operates as follows:\\ 
    \draftonly{\begin{enumerate}
        \item} \dcout{On receiving a request $r$ from an updated client, the manager sends the request to an updated worker, as defined in Algorithm~\ref{alg:ordered}. }
        \draftonly{\item} On receiving a request $r$ from a non-updated client, the update manager checks if $r$ is forward-compatible under a given translation $F$.
        If it is, then the update manager may assign $F(r)$ (treated atomically) to an old worker process or $r$ to a new worker process. 
        Otherwise, $r$ must be assigned to a new version worker process.
    \draftonly{\end{enumerate}}
\end{enumerate}
\end{algorithm}

\subsubsection{Compatibility Rewrites}
\label{subsec:compatible-rewrites}
The proof of update consistency for \cref{alg:backward-compatible} follows a similar structure to the proof for \cref{alg:basic-comm}. The proof requires an additional assumption: that the database behavior depends only on the action and not the identity of the requesting worker process. We say that the database is ``oblivious to worker identity.''

\begin{theorem} \label{thm:backward-compatible-rewrite}
    \Cref{alg:backward-compatible} guarantees a consistent update. 
\end{theorem}
\begin{proof}
    We will establish this through a rewrite strategy \dc{$(\rewriteinv,\rewrite)$ defined as follows.} 
    \dc{The rewrite invariant $\rewriteinv$ consists of all po-computations $\poexec$ such that if there is an updated database transaction $\delta_b$ in the relay of client request $r$ which precedes a non-updated database action, then $\delta_b$ must be servicing $r$'s backward-compatible translation $B(r)$ rather than $r$ itself.  
    Consider a po-computation $\poexec$ controlled by the algorithm. Step \ref{algstep:backward-compatible} of the algorithm enforces that every updated database transaction services a backward-compatible translation $B(r)$ of the sent client request $r$ (which is treated atomically). Hence, $\poexec \in \rewriteinv$.}

    \dc{The po-computation transformation $\rewrite$ is then defined as follows. Consider a computation $\poexec \in \rewriteinv$ and a client $\client{}$.}
    If \dcout{the progress measure is not $0$ in either component,}
    \dc{neither component of the progress measure is 0, then there must be some} \dcout{is an} updated database transaction $\delta_b$ in $\poexec$ which precedes some non-updated database transaction $\delta_r$ \dc{(contributing towards a non-zero sort measure).
    By $\poexec \in \rewriteinv$, $\delta_b$ is servicing} a backward-compatible translation $B(r)$ of \dc{the sent client} \dcout{some} request $r$ (which is treated atomically).
    Let $\worker{u}$ denote the (updated) worker who services $B(r)$ and let $s$ denote the global state in which $\worker{u}$ services $B(r)$.
    The rewrite \dc{$\rewrite_{\client{}}(\poexec)$} introduces a fresh non-updated worker $\worker{n}$ which services request $r$ in global state $\gstate$, and $\worker{u}$ does not service $B(r)$.
    (In other words, we replace $B(r)$ by $\worker{u}$ with $r$ by $\worker{n}$.) This is the only request that $\worker{n}$ services. 
    Let $\delta_t$ denote the non-updated database transaction which services $r$.

    \knout{By our architectural assumptions on the database process, there does not exist any action $\action$ in $\poexec$ such that $\delta_{b,1} < \action < \delta_{b,2}$ for any non-trivial split $\delta_{b,1}\delta_{b,2} = \delta_b$ (and similarly for $\delta_t$ and $\delta_r$). For the remainder of this proof, we therefore treat $\delta_b$, $\delta_t$, and $\delta_r$ as if they were atomic (ordered) actions $\mathit{bd}, \mathit{t}$, and $\mathit{rd}$, respectively. The technical details of the rewrite's replacement of ordered actions in $\poexec$ are similar to what is described in the proof of \cref{thm:single-comm-rewrite}.}
    \kn{As in the proof of \cref{thm:single-comm-rewrite}, we treat $\delta_b$, $\delta_t$, and $\delta_r$ as if they were atomic (ordered) actions $\mathit{bd}, \mathit{t}$, and $\mathit{rd}$, respectively.}

    We need to show that this replacement (I) produces a valid $\client{}$-equivalent po-computation, (II) strictly decreases the progress measure, and (III) \dc{preserves the rewrite invariant $\rewriteinv$} \dcout{property that updated actions preceding non-updated actions are backward-compatible translations of requests}. The claim then follows from \cref{thm:c-rewrite-strategy-update-consistent}. We tackle the required conditions in reverse order.

    \textbf{III. \dc{Preservation of $\rewriteinv$}\dcout{Updated actions preceding non-updated actions are backward-compatible}.} \dc{Backward-compatibility is a property of actions and does not depend on a specific computation. Moreover, the rewrite only decreases the set of updated actions servicing backward-compatible translations. Hence, $\rewriteinv$ is preserved.} \dcout{This condition is a global property of the computation, which is unaffected by the replacement of $bd$ with $t$.}

    \textbf{II. Progress measure decreases.} By Lemma \ref{lem:red-inside-global-update-cut}, every red action is inside the global update cut. 
    Since $bd < rd$ by assumption, $bd$ is inside the global update cut as well, so the cut measure is greater than 0. The substitution replaces (updated) $bd$ with (non-updated) $t$.
    This removes one blue action from the global update cut, decrementing the cut measure and hence decrementing the computation's progress measure.

    \textbf{I. Produces valid $\client{}$-equivalent po-computation.} By our assumption that clients do not know the set of active workers $\workers$, the rewrite can introduce a new, non-updated worker $\worker{n}$. 
    The result of worker $\worker{n}$ servicing the relay produces the same client projections as the original computation for all clients by definition of backward-compatibility, the obliviousness of the database to worker identity, and the fact that clients cannot tell the identities of workers apart. 
    So, this substitution is client-equivalent to the original for all clients.
    
    By definition of backward-compatibility, the final global state resulting from the substituted computation will be equal to the final global state in the original computation.
\end{proof}

The proof for update consistency of \cref{alg:forward-compatible} is analogous.

\subsection{Commutativity + Forward-Compatibility Algorithm}
\label{subsec:comm-forward-comp-alg}

We can combine the notions of commutativity and compatibility to derive additional consistent update algorithms.
We present \cref{alg:comm-update-inv} which exploits commutativity and forward-compatibility; conditions can be flipped for backward-compatibility (see \appendixRef{appendix:compat-update-algorithms}).  
\dc{Orderedness is enforced in a limited way: when worker version can be obscured by forward-compatibility, orderedness is not enforced. The (translation of) the request can be serviced by a non-updated worker even for an updated client, since the client will perceive that worker as updated. However, an updated client must be serviced strictly by updated workers for requests that do not have compatible translations.}

\begin{algorithm}[t]
\caption{Commutativity with Forward-Compatibility\dcout{algorithm} (\dc{Limited} Extension of \cref{alg:ordered})}
\label{alg:comm-update-inv}
\begin{enumerate}
    \item \label{algstep:tell-db-updating-2} At the start, the update manager notifies the database that the update has begun.
    \item \label{algstep:enforce-req-order-2} During the update, the update manager tags all incoming client requests with a number. The database enforces that requests are completed in the assigned order.
    \item The update manager operates as follows:
    \begin{enumerate}
        \draftonly{\item} \dcout{Upon receiving a request $r$ from an updated client, the manager sends the request to an updated worker, as defined in Algorithm~\ref{alg:ordered}.}
        \item \label{algstep:update-inv}  Upon receiving a request $r$\dcout{ from a non-updated client}, the update manager checks if it is forward-compatible under a given translation $F$.
        If it is forward-compatible, then the update manager may assign $F(r)$ (treated atomically) to an old worker process, or $r$ to a new worker process. 
        \item If it is not forward-compatible: 
        \begin{enumerate}
            \item \dc{If $r$ is from an updated client, the manager sends the request to an updated worker, as defined in \cref{alg:ordered}.}
            \item \dc{If $r$ is from a non-updated client,} the update manager checks if $r$ commutes to the left over all prior new-version requests and old-version forward-compatible translations.
            \item \label{algstep:update-vis-left-comm} If $r$ does left-commute, the update manager may assign $r$ to any worker, regardless of version. 
            \item Otherwise, the update manager must assign $r$ to a new version worker process.
        \end{enumerate}
    \end{enumerate}
\end{enumerate}
\end{algorithm}

\begin{theorem}
    \Cref{alg:comm-update-inv} guarantees a consistent update.
\end{theorem}

Intuitively, \dc{the rewrite invariant $\rewriteinv$ includes all po-computations such}
\dcout{Algorithm \ref{alg:comm-update-inv} enforces} that all non-updated jobs are either forward-compatible or left-commutative, \dc{a property enforced by \cref{alg:comm-update-inv}}. \dc{This property ensures} that non-updated jobs can be perceived as either updated or having occurred earlier, before the first perceived update point. The rewrite strategy \dc{$\rewrite$} first replaces a forward-compatibility translation $F(r)$ serviced by a non-updated worker with $r$ serviced by an updated worker, and then left-commutes subsequent non-updated database actions past prior updated database actions (including the database actions servicing the newly introduced $r$).

Note that this rewrite strategy requires a combination of commutativity and compatibility. However, isolating the steps may not lead to a rewrite strategy. Replacing $F(r)$ by an old worker with $r$ by a new worker may not make progress if the database transaction of $F(r)$ is not immediately preceding the final (non-updated) database transaction inside the cut. And the ability to left-commutate an old-version database transaction past prior new-version database transactions is only guaranteed in combination with the compatibility replacement step.

\section{Related Work}
\label{sec:related-work}

\dcout{The question of on-the-fly (also known as ``dynamic'') updates has been investigated in  different settings. We discuss closely related work.}

\smartparagraph{\dc{Type Safety and Invariance Preservation}}
The seminal research on the correctness of dynamic \dc{(on-the-fly)} software update includes that on the Argus system~\cite{bloom-thesis-1983,DBLP:journals/iee/BloomD93} and DYMOS~\cite{DBLP:journals/sigplan/CookL83}. These systems allow the type-safe replacement of a program module that is in a quiescent state, typically under severe constraints to prevent inconsistencies. The work on Argus, for instance, requires that the new module type must be compatible with the old type, in that the new module has the same  future behavior as the module it replaces. (E.g., the new module code may include performance improvements but keep the original interface.) 
Another early system is CONIC~\cite{DBLP:journals/tse/KramerM90}. It defines a small set of management commands (e.g., passivate, activate, unlink, link), using which one can program updates to a transaction-based distributed system. 
The correctness guarantee for the update is in terms of preserving a global invariant.

While invariant or type preservation provides a measure of consistency, such guarantees are necessarily partial in that they do not cover the full behavior of a system; for instance, nothing is said about liveness requirements, such as requiring a proper response to pending requests. Moreover, these internal guarantees do not necessarily imply a consistent view of the update for external clients. A practical difficulty is that few real-world systems have associated formal proofs of invariance (or, more generally, safety); thus, it is a challenge to establish the preservation of type or invariance safety properties for an update.

Language-specific formulations of update consistency include methods such as DSU (dynamic software updates), which are typically for single-server deployments rather than the distributed multi-server setting discussed above and targeted in this paper\dcout{\mbox{~\cite{DSU:conf/pldi/HicksMN,mutatis-mutandis:conf/toplas/StoyleHBSN,Ginseng:conf/pldi/NeamtiuHSO}}}. 
\dc{\citet{DSU:conf/pldi/HicksMN,mutatis-mutandis:conf/toplas/StoyleHBSN,Ginseng:conf/pldi/NeamtiuHSO} develop a calculus and implementation for type-preservation in DSU of C programs, which ensures that any type that is updated is not used concretely after its update point.}
While helpful, type-consistency is a  weak  requirement, as it does not prevent clients from observing responses that may only arise from mixed-mode operation during an update. The work in~\cite{co-specs-DSU:conf/vstte/HaydenMHFF} tackles this concern by requiring the correctness definition to be system-specific and by defining a program transformation that merges the old and new versions to model DSU in a manner that suffices for verification. This method and others that rely on verification are often unworkable in practice simply because many services do not have a precise formal specification, the implementation ranges over multiple programming languages, and verification of such large and complex service implementations is often infeasible in practice. 

\smartparagraph{\dc{Consistency via Resource Replication}}
The blue/green and big-flip update mechanisms are folklore methods~\dc{\cite{blue-green-martin-fowler,brewer-giant-scale-services}}. Consistency comes at the price of significant inefficiency: a blue/green update requires doubling existing compute and memory resources, while a big-flip update halves throughput during the update. The Imago system~\cite{imago:conf/middleware/DumitraN} follows an approach related to the blue/green method. A full system replica is created, then persistent data is transferred from the original version to the updated replica opportunistically. Once all persistent data has been transferred, traffic is switched to the updated replica. The approach taken in \cite{modular-upgrades-ds:conf/ecoop/AjmaniLS} is similar, in that it allows multiple versions of each component to co-exist concurrently during the update. Concurrently active versions of a component must maintain consistency between their internal states; this is handled by defining a state transformation function or an inter-state consistency relation. It might be necessary to drop an incoming request if it cannot be handled consistently by the multiple active versions of an object. Thus, besides requiring additional memory and compute resources, this method may also drop a request at some intermediate point in a chain of related requests. That makes it challenging to preserve global consistency; indeed, the work does not include a formal consistency argument. 

\smartparagraph{\dc{Consistent Network Updates}}
Message drops are less of a concern for updates to network configurations, as network protocols are designed to be resilient to packet drops. A strong consistency criterion for network updates, called \emph{per packet consistency}, is introduced in~\cite{abs-network-updates:conf/sigcomm/ReitblattFRSW}. This property requires that the route taken by a packet through the network during an update either entirely follows the old configuration or entirely follows the new configuration, never a mix of the two. The paper proposes a two-phase algorithm, where old and new configurations co-exist at each router, with new traffic routed only through the new configurations until all old traffic has been processed. This scheme doubles the memory requirements per router. A recently developed method, called ``causal update''~\cite{in-place-network-updates:conf/sigcomm/NamjoshiGS}, performs a network configuration update without additional memory; however, this in-place update process requires packets sent from an old-configuration router to a new-configuration one to be dropped to avoid violating the per-packet consistency requirement.

The causal update algorithm guarantees that the network update appears atomic which, in turn, establishes per-packet consistency. (Per-packet consistency allows one packet from a client to be processed with the new configuration while the next packet from the same client is processed by the old configuration. This is disallowed in the consistency formulation of ~\cite{in-place-network-updates:conf/sigcomm/NamjoshiGS}.) Our definition of update consistency also requires a service update to appear atomic but takes a client-based view, allowing different clients to view the service update as having taken place at different points. 

\smartparagraph{\dc{Analysis Tools}}
Short of guaranteeing update consistency, tools such as monitoring systems, testing systems, and static analyzers can be used to minimize damage from update-induced failures. Gandalf \cite{Gandalf:cnf/usenix/LiCHDHSYLWLC} is an analysis tool for large-scale systems to catch anomalies or bugs during rollouts of software updates, currently used by Microsoft Azure. DUPTester is a testing framework designed to test upgrades on a small scale so upgrade bugs can be found before deployment; DUPChecker is an accompanying static analysis tool that searches for data syntax incompatibilities \cite{survey:conf/sosp/ZhangYJSRLY}. These tools and others can assist in mitigating update failures; we hope that our work can pave the way for provably consistent updates.

\smartparagraph{\dc{Conclusion}}
The consistency formulation in this paper overcomes key disadvantages of prior methods. Consistency is based on the external behavior of the underlying service as seen by its clients, a view more appropriate for a service. The formulation side-steps the difficulty with the lack of formal service specifications in practice by requiring that each client views an update as being atomic. Thus, system designers have only to check (or enforce) that the desired (possibly informal) specifications are maintained by an atomic update, a simpler and more natural task. The formulation and consistency analysis are independent of the programming languages used to implement a system, and are thus more widely applicable. The proposed update mechanisms work in-place, thus avoiding the disadvantages of the big-flip and blue/green methods. Our algorithms do depend on an analysis of commutativity and other properties of the service actions across versions. While we show that this is unavoidable in general, one does need to perform such analysis. Automated commutativity reasoning for programs is quite challenging (cf.~\cite{DBLP:conf/vmcai/PincusK25}), but it appears to be easier to reason about commutativity at the level of the abstract semantics of service operations, as is done in the (informally analyzed) examples in this paper.

\kn{In future work, we plan to take on implementation and optimization concerns, applying our foundations to obtain efficient update algorithms for complex systems.}
\dc{In this vein,} an important direction for further research is to develop consistent update methods for \knout{complex} services that are designed as a network of microservices: that can perhaps be done through a combination of the single-microservice update methods with the network-level causal update mechanism discussed above. The design of efficient, in-place, provably consistent update methods is, we believe, a largely unexplored topic, one with both substantial practical importance and a rich mathematical structure.

\begin{acks}
  This work is supported in parts by %
  the \grantsponsor{GS100000001}{National Science Foundation}{http://dx.doi.org/10.13039/100000001} under the grant agreement~\grantnum{GS100000001}{2304758}.
  We thank our anonymous reviewers for their valuable feedback. We thank Aurojit Panda, Sahil Tikale, and Amit Lieber for providing systems and industry perspectives, and Sahil for suggesting the accept-higher-numbers optimization for \cref{alg:basic-comm}. We also acknowledge many helpful discussions with colleagues at Nokia Bell Labs and New York University.
\end{acks}

\bibliography{references}

@misc{chaitroth2025consistentupdatesscalablemicroservices,
      title={Consistent Updates for Scalable Microservices}, 
      author={Devora Chait-Roth and Kedar S. Namjoshi and Thomas Wies},
      year={2025},
      eprint={2508.04829},
      archivePrefix={arXiv},
      primaryClass={cs.PL},
      url={https://arxiv.org/abs/2508.04829}, 
}

@inproceedings{survey:conf/sosp/ZhangYJSRLY,
author = {Zhang, Yongle and Yang, Junwen and Jin, Zhuqi and Sethi, Utsav and Rodrigues, Kirk and Lu, Shan and Yuan, Ding},
title = {Understanding and Detecting Software Upgrade Failures in Distributed Systems},
year = {2021},
publisher = {Association for Computing Machinery},
address = {New York, NY, USA},
booktitle = {Proceedings of the ACM SIGOPS 28th Symposium on Operating Systems Principles},
pages = {116–131},
numpages = {16},
location = {Virtual Event, Germany},
series = {SOSP '21},
doi = {10.1145/3477132.3483577}
}

@inproceedings{modular-upgrades-ds:conf/ecoop/AjmaniLS,
    author = {Sameer Ajmani and
              Barbara Liskov and
              Liuba Shrira},
   title="Modular Software Upgrades for Distributed Systems",
booktitle="ECOOP 2006 -- Object-Oriented Programming",
year="2006",
publisher="Springer Berlin Heidelberg",
address="Berlin, Heidelberg",
pages="452--476",
    series = "Lecture Notes in Computer Science",
    volume = "4067",
doi = "10.1007/11785477_26"
}

@InProceedings{imago:conf/middleware/DumitraN,
author="Dumitra{\c{s}}, Tudor
and Narasimhan, Priya",
title="Why Do Upgrades Fail and What Can We Do about It?",
booktitle="Middleware 2009",
year="2009",
publisher="Springer Berlin Heidelberg",
address="Berlin, Heidelberg",
pages="349--372",
series="Lecture Notes in Computer Science",
volume="5896",
doi={10.1007/978-3-642-10445-9_18}
}

@inproceedings {Gandalf:cnf/usenix/LiCHDHSYLWLC,
author = {Ze Li and Qian Cheng and Ken Hsieh and Yingnong Dang and Peng Huang and Pankaj Singh and Xinsheng Yang and Qingwei Lin and Youjiang Wu and Sebastien Levy and Murali Chintalapati},
title = {Gandalf: An Intelligent, {End-To-End} Analytics Service for Safe Deployment in {Large-Scale} Cloud Infrastructure },
booktitle = {17th USENIX Symposium on Networked Systems Design and Implementation (NSDI 20)},
year = {2020},
isbn = {978-1-939133-13-7},
address = {Santa Clara, CA},
pages = {389--402},
url = {https://www.usenix.org/conference/nsdi20/presentation/li},
publisher = {USENIX Association},
month = feb
}

@inproceedings{DSU:conf/pldi/HicksMN,
author = {Hicks, Michael and Moore, Jonathan T. and Nettles, Scott},
title = {Dynamic software updating},
year = {2001},
publisher = {Association for Computing Machinery},
address = {New York, NY, USA},
booktitle = {Proceedings of the ACM SIGPLAN 2001 Conference on Programming Language Design and Implementation},
pages = {13–23},
location = {Snowbird, Utah, USA},
series = {PLDI '01},
doi={10.1145/378795.378798}
}

@inproceedings{Ginseng:conf/pldi/NeamtiuHSO,
  author = {Iulian Neamtiu and Michael Hicks and Gareth Stoyle and Manuel Oriol},
  title = {Practical Dynamic Software Updating for {C}},
  booktitle = {Proceedings of the {ACM} Conference on Programming Language Design and Implementation (PLDI)},
  pages = {72--83},
  location = {Ottawa, Canada},
  month = {June},
  year = {2006},
doi={10.1145/1133255.1133991}
}

@article{mutatis-mutandis:conf/toplas/StoyleHBSN,
author = {Stoyle, Gareth and Hicks, Michael and Bierman, Gavin and Sewell, Peter and Neamtiu, Iulian},
title = {Mutatis Mutandis: Safe and predictable dynamic software updating},
year = {2007},
issue_date = {August 2007},
publisher = {Association for Computing Machinery},
address = {New York, NY, USA},
volume = {29},
number = {4},
issn = {0164-0925},
url = {https://doi.org/10.1145/1255450.1255455},
doi = {10.1145/1255450.1255455},
journal = {ACM Trans. Program. Lang. Syst.},
month = {aug},
pages = {22–es},
numpages = {70}
}

@InProceedings{co-specs-DSU:conf/vstte/HaydenMHFF,
author="Hayden, Christopher M.
and Magill, Stephen
and Hicks, Michael
and Foster, Nate
and Foster, Jeffrey S.",
title="Specifying and Verifying the Correctness of Dynamic Software Updates",
booktitle="Verified Software: Theories, Tools, Experiments",
year="2012",
publisher="Springer Berlin Heidelberg",
address="Berlin, Heidelberg",
pages="278--293",
series="Lecture Notes in Computer Science",
volume="7152",
doi="10.1007/978-3-642-27705-4_22"
}

@inproceedings{abs-network-updates:conf/sigcomm/ReitblattFRSW,
author = {Reitblatt, Mark and Foster, Nate and Rexford, Jennifer and Schlesinger, Cole and Walker, David},
title = {Abstractions for network update},
year = {2012},
publisher = {Association for Computing Machinery},
address = {New York, NY, USA},
pages = {323–334},
numpages = {12},
location = {Helsinki, Finland},
series = {SIGCOMM '12},
booktitle = {Proceedings of the ACM SIGCOMM 2012 Conference},
doi={10.1145/2342356.2342427}
}

@inproceedings{in-place-network-updates:conf/sigcomm/NamjoshiGS,
author = {Namjoshi, Kedar S. and Gheissi, Sougol and Sabnani, Krishan},
title = {Algorithms for In-Place, Consistent Network Update},
year = {2024},
publisher = {Association for Computing Machinery},
address = {New York, NY, USA},
booktitle = {Proceedings of the ACM SIGCOMM 2024 Conference},
pages = {244–257},
numpages = {14},
location = {Sydney, NSW, Australia},
series = {ACM SIGCOMM '24},
doi={10.1145/3651890.3672266}
}

@misc{knight-capital-sec,
title = {{K}night {C}apital {LLC} {SEC} {A}dministrative {P}roceedings},
howpublished = {\url{https://www.sec.gov/files/litigation/admin/2013/34-70694.pdf}},
note = {Accessed: 2024-10-16},
year={2013},
key={Knight Capital}
}

@misc{blue-green-martin-fowler,
author={Fowler, Martin},
title = {{B}lue {G}reen {D}eployment},
howpublished = {\url{https://martinfowler.com/bliki/BlueGreenDeployment.html}},
note = {Accessed: 2025-09-15},
year={2010}
}

@article{brewer-giant-scale-services,
author = {Brewer, E.A.},
address = {Los Alamitos},
copyright = {Copyright IEEE Computer Society Jul 2001},
issn = {1089-7801},
journal = {IEEE internet computing},
language = {eng},
pages = {46-55},
publisher = {IEEE},
title = {Lessons from giant-scale services},
volume = {5},
year = {2001},
number = {4},
doi={10.1109/4236.939450}
}

@article{lamport-time-clocks,
author = {Lamport, Leslie},
title = {Time, clocks, and the ordering of events in a distributed system},
year = {1978},
issue_date = {July 1978},
publisher = {Association for Computing Machinery},
address = {New York, NY, USA},
volume = {21},
number = {7},
issn = {0001-0782},
url = {https://doi.org/10.1145/359545.359563},
doi = {10.1145/359545.359563},
journal = {Commun. ACM},
month = jul,
pages = {558–565},
numpages = {8}
}

@inbook{consistent-global-states,
author = {Babao\u{g}lu, \"{O}zalp and Marzullo, Keith},
title = {Consistent global states of distributed systems: fundamental concepts and mechanisms},
year = {1993},
isbn = {0201624273},
publisher = {ACM Press/Addison-Wesley Publishing Co.},
address = {USA},
booktitle = {Distributed Systems (2nd Ed.)},
pages = {55–96},
numpages = {42}
}

@article{chandy-lamport,
author = {Chandy, K. Mani and Lamport, Leslie},
title = {Distributed snapshots: determining global states of distributed systems},
year = {1985},
issue_date = {Feb. 1985},
publisher = {Association for Computing Machinery},
address = {New York, NY, USA},
volume = {3},
number = {1},
issn = {0734-2071},
url = {https://doi.org/10.1145/214451.214456},
doi = {10.1145/214451.214456},
journal = {ACM Trans. Comput. Syst.},
month = feb,
pages = {63–75},
numpages = {13}
}

@inproceedings{mattern1989virtual,
  author = {Mattern, Friedemann},
  booktitle = {Parallel and Distributed Algorithms},
  pages = {215--226},
  publisher = {North-Holland},
  title = {Virtual Time and Global States of Distributed Systems},
  year = 1989
}

@PhdThesis{bloom-thesis-1983,
  author = 	 {Toby Bloom},
  title = 	 {Dynamic Module Replacement in a Distributed Programming System},
  school = 	 {MIT},
  year = 	 1983,
  address = 	 {https://dspace.mit.edu/handle/1721.1/15685}}

@article{DBLP:journals/iee/BloomD93,
  author       = {Toby Bloom and
                  Mark Day},
  title        = {Reconfiguration and module replacement in Argus: theory and practice},
  journal      = {Softw. Eng. J.},
  volume       = {8},
  number       = {2},
  pages        = {102--108},
  year         = {1993}
}

@inproceedings{DBLP:journals/sigplan/CookL83,
  author       = {Robert P. Cook and
                  Insup Lee},
  title        = {{DYMOS:} a dynamic modification system},
  booktitle    = {{SIGSOFT}},
  pages        = {201--202},
  publisher    = {{ACM}},
  year         = {1983},
doi = {10.1145/1006140.1006188}
}

@article{DBLP:journals/tse/KramerM90,
  author       = {Jeff Kramer and
                  Jeff Magee},
  title        = {The Evolving Philosophers Problem: Dynamic Change Management},
  journal      = {{IEEE} Trans. Software Eng.},
  volume       = {16},
  number       = {11},
  pages        = {1293--1306},
  year         = {1990},
doi = {10.1109/32.60317}
}

@inproceedings{DBLP:conf/vmcai/PincusK25,
  author       = {Jared Pincus and
                  Eric Koskinen},
  title        = {An Abstract Domain for Heap Commutativity},
  booktitle    = {{VMCAI} {(2)}},
  series       = {Lecture Notes in Computer Science},
  volume       = {15530},
  pages        = {26--49},
  publisher    = {Springer},
  year         = {2025},
doi = {10.1007/978-3-031-82703-7_2}
}

@article{veracity,
author = {Chen, Adam and Fathololumi, Parisa and Koskinen, Eric and Pincus, Jared},
title = {Veracity: declarative multicore programming with commutativity},
year = {2022},
issue_date = {October 2022},
publisher = {Association for Computing Machinery},
address = {New York, NY, USA},
volume = {6},
number = {OOPSLA2},
url = {https://doi.org/10.1145/3563349},
doi = {10.1145/3563349},
journal = {Proc. ACM Program. Lang.},
month = oct,
articleno = {186},
numpages = {31},
}

@InProceedings{auto-comm-cond,
author="Bansal, Kshitij
and Koskinen, Eric
and Tripp, Omer",
editor="Beyer, Dirk
and Huisman, Marieke",
title="Automatic Generation of Precise and Useful Commutativity Conditions",
booktitle="Tools and Algorithms for the Construction and Analysis of Systems",
year="2018",
publisher="Springer International Publishing",
address="Cham",
pages="115--132",
isbn="978-3-319-89960-2",
doi="10.1007/978-3-319-89960-2_7"
}

@ARTICLE{UpdateSurvey,
  author={Foerster, Klaus-Tycho and Schmid, Stefan and Vissicchio, Stefano},
  journal={IEEE Communications Surveys \& Tutorials}, 
  title={Survey of Consistent Software-Defined Network Updates}, 
  year={2019},
doi={10.1109/COMST.2018.2876749}
}

\iftoggle{techrep}{
  \newpage
  \appendix
  \section{Appendix}
\label{sec:appendix}

\subsection{Lemmas for Section \ref{sec:formal-model}}
\label{appendix:formal-model}

The following lemmas state properties of the architecture that are needed for the consistency proofs. 
\begin{lemma}
\label{lem:action-in-relay}
    Every action in a microservice is part of exactly one relay.
\end{lemma}

\begin{lemma}
\label{lem:relay-procs}
    Every relay includes exactly one client, exactly one worker, and the database.
\end{lemma}

\subsection{Theorems and Proofs for Section \ref{sec:proving-update-consistency}}
\label{appendix:proving-update-consistency}

\begin{proof}[Proof for Lemma~\ref{lem:msg-pass-bet-procs}]
    Straightforward from the happens-before relation.
\end{proof}

\begin{proof}[Proof for Lemma~\ref{lem:red-inside-global-update-cut}]
    Suppose to the contrary, that there is a red action outside the global update cut. Call this action $r$.
    Since $r$ is outside the cut, it must not be in the downwards closure of any red client response receipt.
    By Lemma \ref{lem:action-in-relay}, $r$ must be part of a relay, and by definition, the client receive action in $r$'s relay must be red.
    But then $r$ is in the downwards closure of a red client response receipt. Contradiction.
\end{proof}

\begin{proof}[Proof for Theorem~\ref{thm:cut-0-atomic}]
  By assumption, there exists a consistent cut of $\poexec$ with no updated actions inside the cut and no non-updated actions outside the cut. By definition, the computation has an update-consistent cut, so by Theorem \ref{thm:update-consistent-cut}, $\poexec$ has an atomic update.
\end{proof}

\begin{proof}[Proof for Lemma~\ref{lem:b-rr-distinct-workers}]
    Since $\mathit{rr}$ is a red response, the worker in $\mathit{rr}$'s relay must be non-updated. Call this worker $\worker{r}$ and the first worker action in the relay $\mathit{rr}'$. Since $b$ is a blue action, the worker in $b$'s relay must be updated. Call this worker $\worker{b}$ and the first worker action in the relay $b'$ (possibly with $b = b'$). 
    
    Suppose $\worker{r} = \worker{b}$. Since workers service only one client and one request at a time, and $b < \mathit{rr}$, we must also have $b' < \mathit{rr}'$. But this would mean that the worker updated and then reverted the update, which is not possible.
\end{proof}

\begin{proof}[Proof for Theorem~\ref{thm:sort-0-consistent-cut}]
    \dc{Assume $\poexec$ has a sort progress measure of 0.}
    We will construct a consistent cut of $\poexec$ that we show to be update-consistent. 
    \dc{By \cref{thm:update-consistent-cut}, $\poexec$ must have an atomic update.}
    Our chosen cut $\cut$ is the global update cut. The global update cut is consistent by construction, so it remains to show that it is update-consistent.
    
    Suppose $\cut$ is not update-consistent. We will prove that this contradicts the sort progress measure being 0.
    
    By Lemma \ref{lem:red-inside-global-update-cut}, there cannot be any red actions outside the cut. Since $\cut$ is not update-consistent, there must be a blue action inside the cut. Let $b$ be such a blue action. Since $b$ is inside the global update cut, it must be in the downwards closure of some red client response receive action. Let $\mathit{rr}$ be such an action, so that we have $b < \mathit{rr}$.

    By Lemma \ref{lem:action-in-relay}, $\mathit{rr}$ is in a red relay, and there must be a red database receive action preceding $\mathit{rr}$ in its relay.\draftonly{\footnote{\dcout{The theorem can be proved without the assumption of a database access in the relay, but it relies on a more complex analysis. The proof relies on the fact that distinct clients and workers only interact at the database.}} }
    Call this database action $\mathit{rd}$. \dc{We will prove that there must exist a blue database operation $\mathit{bd}$ such that $\mathit{bd} < \mathit{rd}$, violating the assumption that the sort progress measure is 0.}

    \dc{First, suppose $b$ and $rr$ service distinct clients.} By Lemma\dcout{s \ref{lem:b-rr-distinct-clients} and} \ref{lem:b-rr-distinct-workers}, $b$ and $\mathit{rr}$ must \dcout{service distinct clients and} be serviced by distinct workers. Distinct clients serviced by distinct workers only interact at the database according to the system architecture. \dcout{We will prove that there must exist a blue database operation $\mathit{bd}$ such that 
    $\mathit{bd} < \mathit{rd}$,
    violating the assumption that the sort progress measure is 0.}

    Suppose $b$ is an action preceding or equal to the database's send-result action in its relay: a client send request, worker receive request, worker send op, database receive op, or database send result. By Lemma \ref{lem:action-in-relay}, $b$ is in a blue relay, and by our assumption, $b \leq \mathit{bd}$, where $\mathit{bd}$ is the (blue) database action in $b$'s relay. 
    If $\mathit{rd} < bd$, then no subsequent database interactions can invert that order, so $b$ and $\mathit{rr}$ would be unrelated. Thus, to have $b < \mathit{rr}$, we must have $\mathit{bd} < \mathit{rd}$. 
    
    Alternatively, suppose $b$ is in the latter half of its relay: a worker receive result, worker send result, or client receive result. By Lemma \ref{lem:action-in-relay} again, $b$ is in a blue relay, and there must be a blue database receive action preceding it in the relay. Call this blue database action $bd$. 
    Since $b < rr$, we must also have $bd < rr$, and since distinct clients and workers only interact at the database, we must then have $bd < rd$.

    \dc{Next, suppose $b$ and $rr$ service the same client (which may occur in a non-ordered update). By \cref{lem:action-in-relay} and \cref{lem:relay-procs}, $b$ and $rr$ must be in distinct relays, each with a database action. Call the database action in $b$'s relay $bd$ and the database action in $rr$'s action $rd$. Since clients are sequential and $b < rr$, we must have $bd < rd$. }
    
    We now have that regardless of what type of action $b$ is, there must exist a blue database action $\mathit{bd}$ such that $\mathit{bd} < \mathit{rd}$. But this violates the assumption that the sort progress measure is 0. 
    We can conclude that the global update cut on the computation is update-consistent, \dc{and thus the update is consistent}.
\end{proof}

\begin{proof}[Proof for Lemma~\ref{lem:b-rr-distinct-clients}]
  If $b$ is performed by the same client $\client{}$ who performs $\mathit{rr}$, then by $b < \mathit{rr}$, $b <_{\client{}} \mathit{rr}$. But this violates orderedness.
  
  Likewise, suppose $b$ is performed by a worker or the database, and $b$ services the same client who performs $\mathit{rr}$. Since $b$ is a blue action, the response generated by $b$ is blue as well. Call this response $br$. Since clients are sequential and $b < \mathit{rr}$, we must also have $br <_{\client{}} \mathit{rr}$. But this again violates orderedness.
\end{proof}

\begin{proof}[Proof Details for Part II.(C) of \cref{thm:single-comm-rewrite}]
    Note that since $\RD < \BD$ in the rewrite, $\BD < r$ in the rewrite means we must have $\BD < r$ in the original computation. There are a few cases to consider.
    
    (1) $y < \BD$ in the original computation. Since $\BD < r$ in the original computation, $y$ is in $C$.
    
    (2) $y < \RD$ in the original computation. First, note that the relation $\RD < \BD$ in the rewrite must be strictly by database process order, since otherwise there would be a loop in the original computation. Now there are two possibilities.
    
    (2a) $\BD < r$ in the original computation is witnessed by a chain that touches the database, i.e. we have $\BD < d < r$ in the original for some database event $d$. This event must then succeed $\RD$ in the original, giving us $\RD < d < r$, and so $y$ is in $C$.
    
    (2b) No chain witnessing $\BD < r$ in the original computation touches the database, excluding $y$ from $C$. This case is not possible. Since this is an ordered update by definition of $\rewriteinv$, we conclude that the relays containing $\BD$ and $r$ belong to distinct clients (by \cref{lem:b-rr-distinct-clients}) and have distinct workers (by \cref{lem:b-rr-distinct-workers}). As clients and workers are mutually independent, the dependency $\BD < r$ must arise through the database.
\end{proof}

\subsection{Additional Algorithm for \cref{sec:compat-update-algorithms}}
\label{appendix:compat-update-algorithms}

\begin{algorithm}
\caption{Commutativity with Backward-Compatibility\dcout{algorithm} (Extension of \cref{alg:ordered})}
\label{alg:comm-backward-compatible}
\begin{enumerate}
    \item \label{algstep:tell-db-updating-3} At the start of the update, the update manager notifies the database that the update has begun.
    \item \label{algstep:enforce-req-order-3} During the update, the update manager tags all incoming client requests with a number. The database enforces that requests are completed in the assigned order.
    \item The update manager operates as follows:
    \begin{enumerate}
        \item Upon receiving a request $r$ from an updated client, the manager sends the request to an updated worker, as defined in Algorithm~\ref{alg:ordered}.
        \item \label{algstep:update-inv-2} Upon receiving a request $r$ from a non-updated client, the update manager checks if $r$ commutes to the left over all prior new-version requests that are not backward-compatible translations.
        If it is left-commutative, the update manager may assign $r$ to any worker, regardless of version. 
        \item If $r$ is not left-commutative, then the update manager checks if $r$ is backward-compatible under a given translation $B$.
        If $r$ is backward-compatible, then the update manager checks if updated $B(r)$ commutes to the left over all prior new-version requests \dc{that are not backward-compatible translations}. 
        If updated $B(r)$ does left-commute, the update manager may assign $B(r)$ or $r$ to an updated worker.
        Otherwise, the update manager must assign $r$ to an updated worker.
    \end{enumerate}
\end{enumerate}
\end{algorithm}

\dc{Intuitively, \cref{alg:comm-backward-compatible} enforces that every non-updated action left-commutes past every prior updated action that is visibily updated (ie, not a backward-compatible translation). Every updated action that appears to be non-updated (ie, is a backward-compatible translation) also left-commutes past every prior visibly updated action.
Orderedness is fully enforced here because we are not manipulating a client's point of update. Each client's update point is fixed when they receive their first updated response. Instead, we are using backward-compatibility to shift which actions must commute to guarantee orderedness.}
}{}

\end{document}